\documentclass[11pt]{article}
\usepackage{fancyhdr,epsfig,lastpage}
\usepackage{amsmath}
\usepackage{latexsym}
\usepackage{url}
\usepackage[colorlinks]{hyperref}
\usepackage[dvips]{color}
\usepackage[numbers,sort&compress]{natbib}
\usepackage{pstricks}
\begin{document}
 \newcommand{\fn}[1]{\footnote{ #1}}
 \newcommand{\mean}[1]{\left\langle #1 \right\rangle}
 \newcommand{\abs}[1]{\left| #1 \right|}
 \newcommand{\la}{\langle}
 \newcommand{\ra}{\rangle}
 \newcommand{\RA}{\Rightarrow}
 \newcommand{\tet}{\vartheta}
 \newcommand{\eps}{\varepsilon}
 \newcommand{\bbox}[1]{\mbox{\boldmath $#1$}}
 \newcommand{\ul}[1]{\underline{#1}}
 \newcommand{\ol}[1]{\overline{#1}}
 \newcommand{\non}{\nonumber \\}
 \newcommand{\no}{\nonumber}
 \newcommand{\eqn}[1]{eq. (\ref{#1})}
 \newcommand{\Eqn}[1]{Eq. (\ref{#1})}
 \newcommand{\eqs}[2]{eqs. (\ref{#1}), (\ref{#2})}
 \newcommand{\pics}[2]{Figs. \ref{#1}, \ref{#2}}
 \newcommand{\pic}[1]{Fig. \ref{#1}}
 \newcommand{\sect}[1]{Sect. \ref{#1}}
 \newcommand{\name}[1]{{\rm #1}}

\newcommand{\rb}[1]{\raisebox{-1 ex}{#1}}
\newcommand{\av}[1]{\left< #1 \right>}
\newcommand{\abst}[0]{\rule[-1.5 ex]{0 ex}{4 ex}}
\newcommand{\fref}[1]{Fig.~\ref{#1}}
\newcommand{\tref}[1]{Tab.~\ref{#1}}
\newcommand{\eref}[1]{Eq.~(\ref{#1})}
\newcommand{\sref}[1]{Section~\ref{#1}}
\newcommand{\aref}[1]{Appendix~\ref{#1}}

 \newcommand{\vol}[1]{{\bf #1}}
 \newcommand{\et}{{\it et al.}}
 \newcommand{\D}{\displaystyle}
 \newcommand{\T}{\textstyle}
 \newcommand{\SC}{\scriptstyle}
 \newcommand{\SSC}{\scriptscriptstyle}
 \renewcommand{\textfraction}{0.01}
 \renewcommand{\topfraction}{0.99}
 \renewcommand{\bottomfraction}{0.99}
 \renewcommand{\floatpagefraction}{0.99}
\renewcommand{\thefootnote}{\fnsymbol{footnote}}

 \begin{center}
   {\Large \bf On Spatial Consensus Formation: \\[1mm]
Is the Sznajd Model Different from a Voter Model?
}\\[5mm]

{ \large \bf Laxmidhar Behera$^{1,2}$, Frank
  Schweitzer$^{1,3,}$\footnote[5]{Corresponding author:
    \url{schweitzer@ais.fraunhofer.de} }
} 
\begin{quote}
\begin{itemize}
\item[$^{1}$]{\it Fraunhofer Institute for Autonomous Intelligent
    Systems, Schloss Birlinghoven, D-53757 Sankt Augustin, Germany}
\item[$^{2}$] \emph{Department of Electrical Engineering, 
Indian Institute of Technology \\ Kanpur 208 016, India} 
\item[$^{3}$]{\it Institute of Physics, Humboldt University Berlin,
    10099 Berlin, Germany}
\end{itemize}
\end{quote}
 \end{center}

\begin{abstract}
  
  In this paper, we investigate the so-called ``Sznajd Model'' (SM) in
  one dimension, which is a simple cellular automata approach to
  consensus formation among two opposite opinions (described by spin up
  or down).  To elucidate the SM dynamics, we first provide results of
  computer simulations for the spatio-temporal evolution of the opinion
  distribution $L(t)$, the evolution of magnetization $m(t)$, the
  distribution of decision times $P(\tau)$ and relaxation times $P(\mu)$.
  
  In the main part of the paper, it is shown that the SM can be
  completely reformulated in terms of a linear VM, where the transition
  rates towards a given opinion are directly proportional to frequency of
  the respective opinion of the second-nearest neighbors (no matter what
  the nearest neighbors are).  So, the SM dynamics can be reduced to one
  rule, ``Just follow your second-nearest neighbor''. The equivalence is
  demonstrated by extensive computer simulations that show the same
  behavior between SM and VM in terms of $L(t)$, $m(t)$, $P(\tau)$,
  $P(\mu)$, and the final attractor statistics.
  
  The reformulation of the SM in terms of a VM involves a new parameter
  $\sigma$, to bias between anti- and ferromagnetic decisions in the case
  of frustration. We show that $\sigma$ plays a crucial role in explaining
  the phase transition observed in SM.  We further explore the role of
  synchronous versus asynchronous update rules on the intermediate
  dynamics and the final attractors. Compared to the original SM, we find
  three additional attractors, two of them related to an asymmetric
  coexistence between the opposite opinions.

\end{abstract}

\section{Introduction}

The old wisdom still holds: if a single person finds a particular case
important, this does not matter too much -- but already if two persons
are convinced of its importance, they have a good chance to convince
others. This can be simulated by means of a cellular automaton (CA) model
of consensus formation, meanwhile well known as \textsl{Sznajd model}
(abbreviated as SM from now on). It is named after the two Polish
authors, Katarzyna Sznajd-Weron and her father J\'ozef Sznajd, who in
2000 published a paper on ``opinion evolution in a closed community''
\citep{weron:2000}. Interestingly, the dynamics of convincing others can
be applied to the adoption of the SM in the scientific community itself.
The paper and the rather simple model (discussed in Sect.  2) would
problably not have gained so much attention without at least one other
person confident about its importance. It was Dietrich Stauffer
\citep{sousa:2000, stauffer-03, stauffer:2001, 
  stauffer-02acs}, who, after being influenced by the positive response
of his collaborators, started to propagate the SM in various scientific
communities: physicists, social scientists, computer scientists -- and
this way persuaded ``neighboring'' scientists \citep{moreira-01,
  chang-01, bernardes:2002, elgazzar:2001, ochrombel:2001,
  bernardes-02ij, newscientist} to play with it (including the authors of
this paper).  So the SM -- which the original authors called \emph{USDF
  model}: ``united we stand, divided we fall'' -- attracted a lot of
interest. In particular, the dynamics, originally given for the
one-dimensional lattice, was generalized to higher dimensions
\citep{sousa:2000, bernardes:2002}.  \citet{bernardes-02ij} used the SM
to explain the distribution of political votes and in
\citep{bernardes:2002} applied it, together with a {B}arabasi network, to
the Brazilian elections.  In \citep{elgazzar:2001}, the SM was also
applied to small-world networks.  Other applications deal with financial
markets \citep{weron:2002}, with aspects of statistical physics, such as
correlated percolation \citep{moreira-01}, and with different geometries
\citep{chang-01}.

While we on one hand are allured by the complex intermediate dynamics of
this rather simplistic model, our interest in this subject is mainly
driven by the question: Is there anything new in the SM? I.e., in what
respect is the SM different from other CA models dealing with local
adoption processes? 

In fact, ever since CA started to invade the social sciences in the
1950's, lots of different CA-based models were proposed to describe
spatial structure formation in social systems \citep{neumann-66,
  sakoda-71, schelling-69, schelling-71, albin-75, hegselmann-flache-98,
  kacp-holyst-96, solomon-et-00}. One well established model class is
known as the \emph{voter model} (abbreviated as VM from now on). It is
based on the idea that the adoption of a given ``opinion'' (behavior,
attitude) depends on the frequency of that opinion in the neighborhood.
In the \emph{linear} VM, the transition rate of adopting an opinion is
directly proportional to the given frequency, in non-linear VM also other
frequency dependencies (e.g. voting against the trend) are taken into
account. VM with positive frequency dependence (i.e.  \emph{majority
  voting rules}) can be considered as a simple prototype for modeling
\emph{herding behavior} -- a phenomenon widely known in biology, economy
and the social sciences. In a special class of non-linear VM, the
non-linearity may result from certain economic or social considerations,
for example from a payoff matrix. Hence, as long as the adoption dynamics
depend on the (local) frequency, even applications in population biology
or evolutionary game theory can be treated as a non-linear VM.

In addition to earlier work on mathematical analysis on VM \citep{lig94},
\emph{spatial} VM have been recently investigated by means of CA concepts
\citep{Durrett:94, Durrett:99, Molofsky:99, Dieckmann:00, Krapivsky:03}.
In our own work \citep{lb-fs-1d-03, fs-voter-03}, we were particularly
interested in spatio-temporal pattern formation in one- and
two-dimensional VM. Based on theoretical investigations of the
microscopic CA dynamics, we were able derive approximations for
time-dependent macroscopic quantities, such as frequencies or spatial
correlations, and critical parameters for phase transitions. A comparison
of these results with the CA simulations showed a good agreement.  Other
investigations of the macroscopic VM dynamics by means of either pair
approximation or Markov models can be found in \citep{Matsuda:87,
  Matsuda:92, Harada:94, Sato:00, Baalen:00}.

Application of the non-linear voter dynamics to the local adoption of
successful strategic behavior (such as to cooperate or to defect)
revealed phase diagrams for specific spatial patterns (formation of small
clusters and spatial domains, front dynamics) \citep{fs-lb-acs-02}.
Similar work on the spatial adoption of game-theoretical strategies has
been done in \citep{Nowak:92, Nowak:93, Lindgren:94, Nakamaru:97}.

So, given the extensive work done on VM, the question addressed in this
paper is about connections between VM and SM. Evidently, both models deal
with \emph{local adoption processes}, but an in-depth analysis of SM is
still lacking. If we could reveal -- as we will do in this paper -- that
SM is just a special case of VM, then many of the techniques and results
obtained earlier for linear and non-linear VM can be applied to SM, thus
providing us with a wealth of analytical understanding of the SM
dynamics. Hence, to appreciate the ideas behind the SM, we should also
understand how the SM is related to the existing classes of CA models on
consensus formation and what the difference are. To find this out, we
start in Sect. 2 with a brief description of the SM and present computer
simulations in one dimension.  In Sect. 3, we reformulate the SM in terms
of a VM, to show that the SM is in fact a \emph{linear} VM. In Sect. 4,
we further explain the nature of the phase transitions observed in SM by
means of an external parameter, $\sigma$.  In Sect. 5 we investigate the
influence of two different update rules (asynchronous vs.  synchronous)
on the intermediate dynamics and the stationary states. We also show some
new results on consensus formation with asymmetric coexistence.  In Sect.
6, we conclude with some hints about further research.

\section{Dynamics of the SM}
\label{2}

\subsection{Rules of the SM Game}
\label{sec:2.1}

In their original paper, Sznajd-Weron and Sznajd \citep{weron:2000}
proposed a one-dimensional Ising spin model with periodic boundary
conditions where each spin (or lattice site) $i=1,...,N$ can have be
found in one of two states, $\theta_{i}\in \{-1,+1\}$ or $\{-,+\}$ for
short, which in the context of opinion formation shall refer to two
opposite opinions (see \pic{ca}).

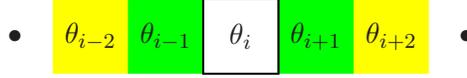
\begin{figure}[htbp]
\vspace*{-1cm}
  \begin{center}
\psset{unit=1.0}
    \begin{pspicture}(7,3)
\rput(0.5,1.5){$\bullet$}
\rput(6.5,1.5){$\bullet$}
\pspolygon[fillcolor=yellow,fillstyle=solid,linestyle=none]
(1,1)(2,1)(3,1)(4,1)(5,1)(6,1)(6,2)(5,2)(4,2)(3,2)(2,2)(1,2)(1,1)
\pspolygon[fillcolor=green,fillstyle=solid,linestyle=none]
(2,1)(3,1)(4,1)(5,1)(5,2)(4,2)(3,2)(2,2)(2,1)
\pspolygon[fillcolor=white,fillstyle=solid,linestyle=solid]
(3,1)(4,1)(4,2)(3,2)
      \rput(3.5,1.5){$\theta_{i}$}
      \rput(2.5,1.5){$\theta_{i-1}$}
     \rput(4.5,1.5){$\theta_{i+1}$}
      \rput(1.5,1.5){$\theta_{i-2}$}
      \rput(5.5,1.5){$\theta_{i+2}$}
\end{pspicture}
\end{center}
\vspace*{-1cm}
\caption{
  One-dimensional CA, where each lattice site $i=1,...,N$ can have one
  out of two opinions, $\theta_{i}\in \{-1,+1\}$ or $\{-,+\}$ for short.
  In the SM, the opinions $\theta_{i-1}$ and $\theta_{i+2}$ depend on the
  opinions of the pair $\theta_{i}$, $\theta_{i+1}$ as described by rules
  1, 2.
\label{ca}}
\end{figure}
For the spin interaction of neighboring spins, the following two rules
are proposed:
\begin{description}
\item[Rule 1:] If two consecutive lattice sites have the \emph{same}
    opinion, (either +1 or -1), i.e. if \mbox{$\theta_{i}
    \theta_{i+1}=1$}, then the two neighbouring sites $\{i-1\}$ and
  $\{i+2\}$ will adopt the opinion of the pair $\{i,i+1\}$, i.e.
  $\theta_{i-1}=\theta_{i+2}=\theta_{i}=\theta_{i+1}$. This rule refers
  to \emph{ferromagnetism}.
\item[Rule 2:] If two consecutive lattice sites have a \emph{different}
  opinion, (either +1 or -1), i.e. if \mbox{$\theta_{i}
    \theta_{i+1}=-1$}, then the two neighbouring sites $\{i-1\}$ and
  $\{i+2\}$ will adopt their opinion from the \emph{second nearest
    neighbors} as follows: $\theta_{i-1}=\theta_{i+1}$,
  $\theta_{i+2}=\theta_{i}$. This rule refers to
  \emph{anti-ferromagnetism}.
\end{description}
A \emph{deterministic dynamics} is considered here, i.e. the rules apply
with probability one -- which is similar to an Ising system at
temperature $T=0$. But there is still randomness in the system in the
sense that (i) there is an inital random distribution of the opinions
with the mean values of the frequencies $f_{+1}=f_{-1}=0.5$, and (ii)
during the computer simulations, the site $i$ for the next step is chosen
randomly, i.e. the dynamics is governed by a  \emph{random sequential
  update}, or \emph{asynchronous} update. 

We further note that in the SM 
\emph{two spins} are flipped at a time. With the two rules above, we find
the following possible transitions in a neighborhood of $n=4$: 
\begin{equation}
  \label{eq:rules}
  \begin{array}{cccccccccc}
\theta_{i-1}& \theta_{i}& \theta_{i+1}& \theta_{i+2} &\quad \to \quad & 
\theta_{i-1}& \theta_{i}& \theta_{i+1}& \theta_{i+2} &\quad \mathrm{rule} 
\quad  \\\hline
? & + &+& ? && +& +& +& + & (1)\\
? & - &-& ? && -& -& -& - & (1)\\
? & + &-& ? && -& +& -& + & (2)\\
? & - &+& ? && +& -& +& - & (2)\\
\end{array}
\end{equation}
 
It has been established through micro-simulations that the
one-dimensional SM for any random initial configuration asymptotically
reaches one of three possible attractors, two of which refer to
ferromagnetism and one to anti-ferromagnetism. These possible attractors
are reached with different probability: 
\begin{itemize}
\item attractor \emph{ferro$_{+}$}: $\{+++++++++\}$ with probability
  $p=0.25$
\item attractor \emph{ferro$_{-}$}: $\{---------\}$ with probability
  $p=0.25$
\item attractor \emph{anti-ferro}: $\{-+-+-+-+-+\}$ 
  with probability $p= 0.5$
\end{itemize}

In order to verify these probabilities, let us consider a lattice of size
$N$ with periodic boundary conditions and an initial random distribution
of $+$ and $-$. Then, the number of consecutive pairs, $\{i,i+1\}$ is
also $N$, and the initial probability of finding either a ferromagnetic
or an anti-ferromagnetic pair adds up to 0.5, i.e.,
\begin{equation}
  \label{anti}
p_{\mathrm{f}}=p_{++}+p_{--}=0.5\;; \quad
p_{\mathrm{af}}=p_{-+}+p_{+-}=0.5\;; \quad 
p_{\mathrm{f}}+p_{\mathrm{af}}=1
\end{equation}
Under these conditions, what is the probability to find ferromagnetic and
anti-ferromagnetic pairs in the course of time? During the first $q$
steps, we may assume that the initial distribution is not changed much by
the dynamics, i.e.  \eqn{anti} remains valid and the probabilities are
given by the binomial distribution:
\begin{equation}
\label{begin}
\sum_{k=0}^q {q \choose k}\, p_{\mathrm{af}}^k \, p_\mathrm{f}^{q-k}=1
\end{equation}
If during the first $q$ steps more than $q/2$ antiferrormagnetic pairs
are selected, then $p_{\mathrm{af}}$ increases since each selection will
lead to two new antiferromagnetic pairs.  The case of ferromagnetic pair
selection can be treated similarly. For $q$ being an even number, the
probability is then given by:
\begin{equation}
\label{eq:prob}
\sum_{i=(q/2)+1}^q {q \choose k} \, 
p_{\mathrm{af}}^k \, p_\mathrm{f}^{q-k}+ \frac{1}{2}
{q \choose q/2}\, p_{\mathrm{af}}^{q/2}\, p_\mathrm{f}^{q/2} =0.5
\end{equation}
where the first term denotes the probability of selecting more than $q/2$
anti-ferromagnetic pairs (favoring anti-ferromagnetism) and the second
term denotes the probability of selecting exactly $q/2$ pairs (favoring
both ferro- and anti-ferromagnetism with probability 0.5). Thus, we can
conclude that the probability for the system to reach the
\emph{anti-ferromagnetic} attractor is given by 0.5, \eqn{eq:prob}. 

Equation \eqref{eq:prob} is valid as long as
$p_\mathrm{f}=p_{\mathrm{af}}=0.5$, i.e. for $t\leq q$. After the initial
time lag, the symmetry is broken and the system dynamics goes towards one
of the possible anti- or ferromagnetic attractors with probability one.

\subsection{Results of SM Computer Simulations}
\label{sec:2.2}

In this section, we show some results of computer simulations of the
one-dimensional CA, using the asynchronous update rule. The results are
basically known, but we present them here to allow for a comparison with
the voter model in the following sections. \pic{fig:Snaid_01} and
\pic{fig:Snaid_05} show the spatio-temporal evolution of the opinion
distribution for the two different attractors and the respective
magnetization curves over time, $m(t)$. The magnetization gives a measure
for reaching the attractor and is defined as:
\begin{equation}
  \label{magnet}
  m(t)=f_{+}-f_{-}\;; \quad
  f_{+}=\frac{1}{N}\sum_{i}^{N}\delta_{+1;\theta_{i}}\;;
\quad  f_{-}=1-f_{+} 
\end{equation}
and $\delta_{+;\theta_{i}}$ is the Kronecker delta, which is 1 only if
$\theta_{i}=+1$ and zero otherwise.
\begin{figure}[htbp]
\vspace*{-10pt}
\centerline{
\psfig{file=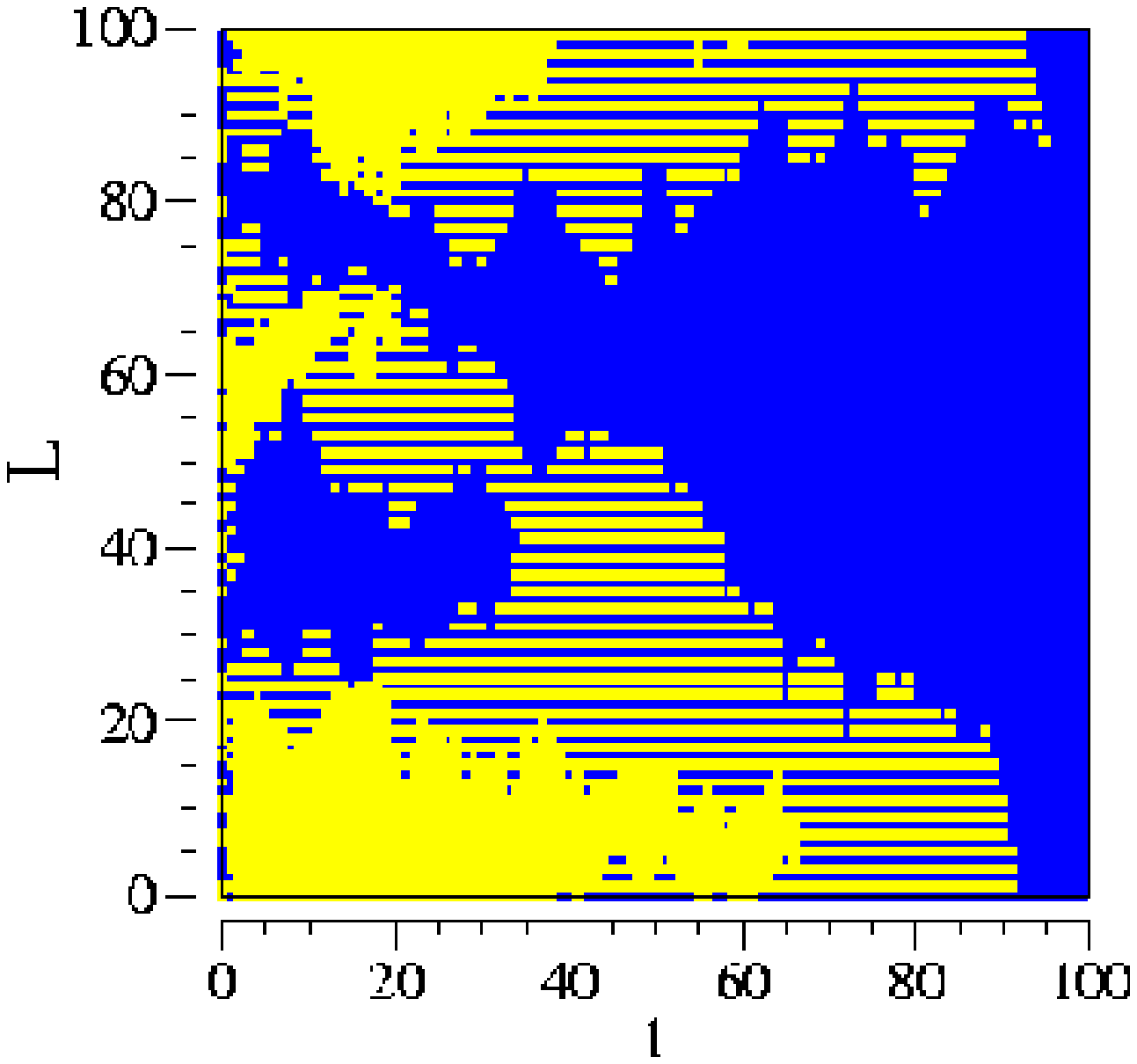,width=8cm}\hfill 
\epsfig{file=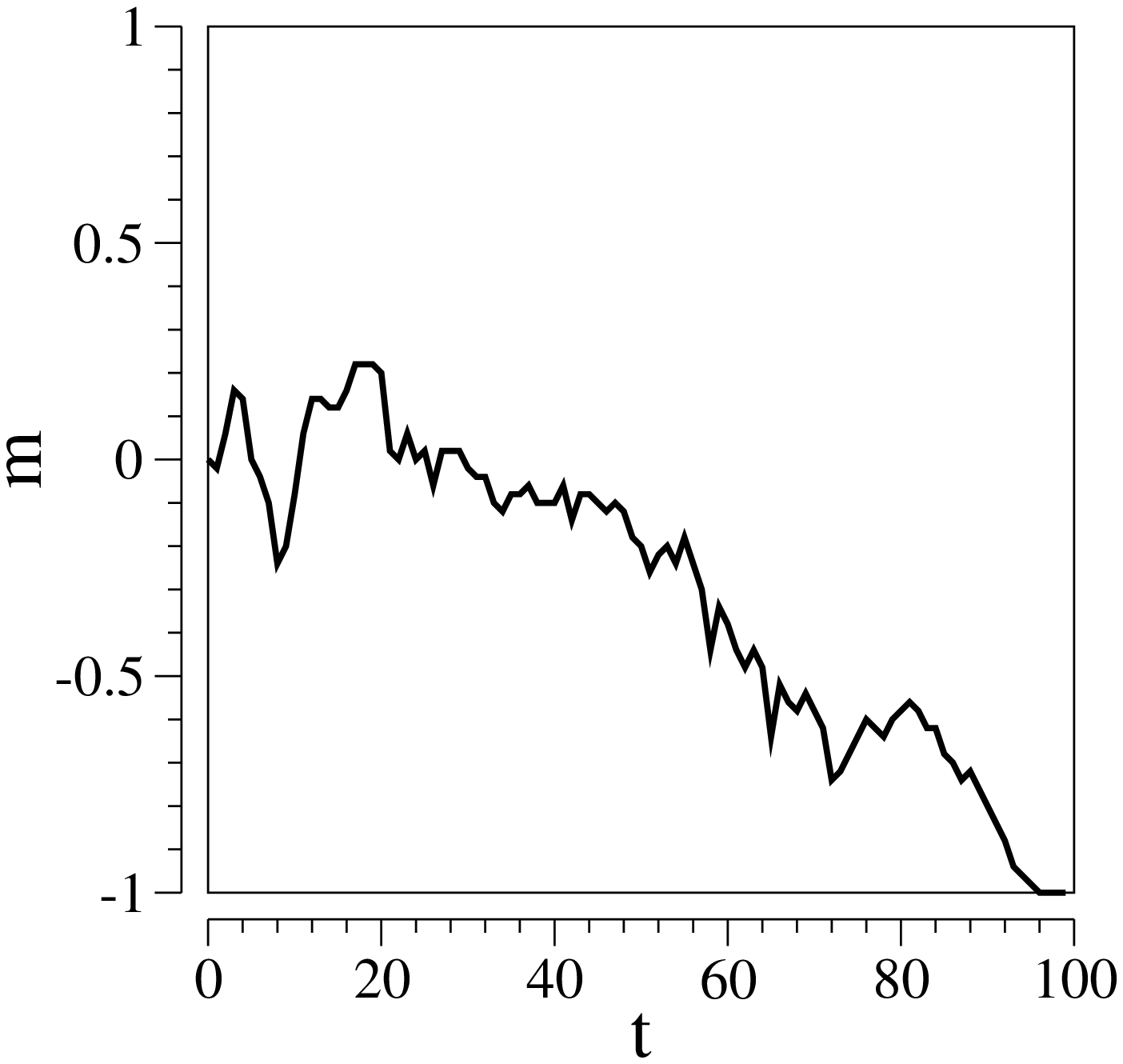,width=8cm} 
}
\vspace*{-10pt}
  \caption{Evolution of the one-dimensional lattice (left) and 
    magnetization $m$ (right) vs time in Monte-Carlo steps.  At
    intermediate times, both anti-ferromagnetic stripes and ferromagnetic
    domains coexist, but asymptotically, the \emph{consensus attractor}
    is reached. ($N=100$, asynchronous update according to the Sznajd
    rules, \eqn{eq:rules}, dark gray dots indicate state $-1$, light gray
    dots state $+1$.) }
  \label{fig:Snaid_01}
\end{figure}

\begin{figure}[htbp]
\vspace*{-10pt}
\centerline{\psfig{file=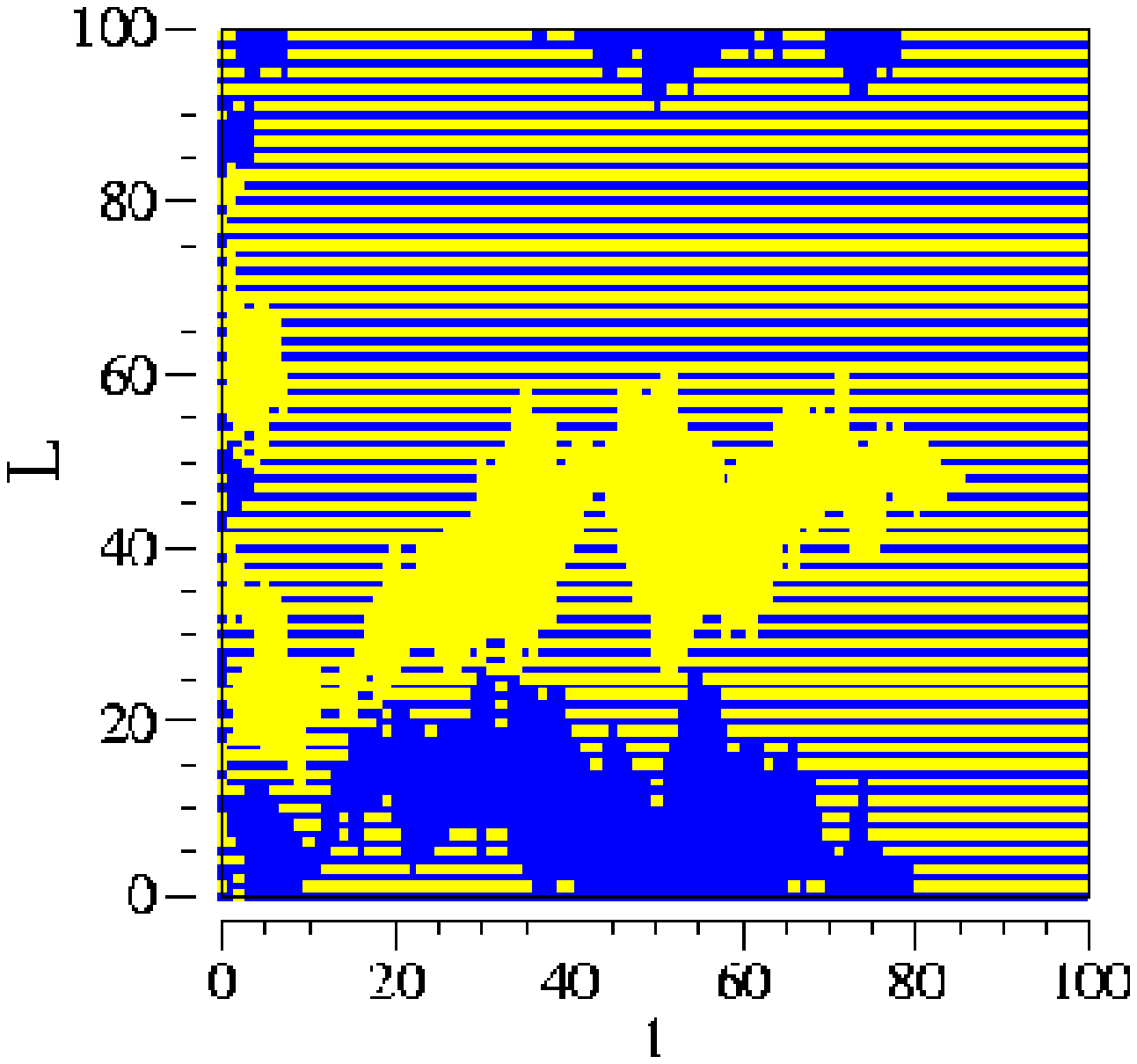,width=8cm} \hfill
\psfig{file=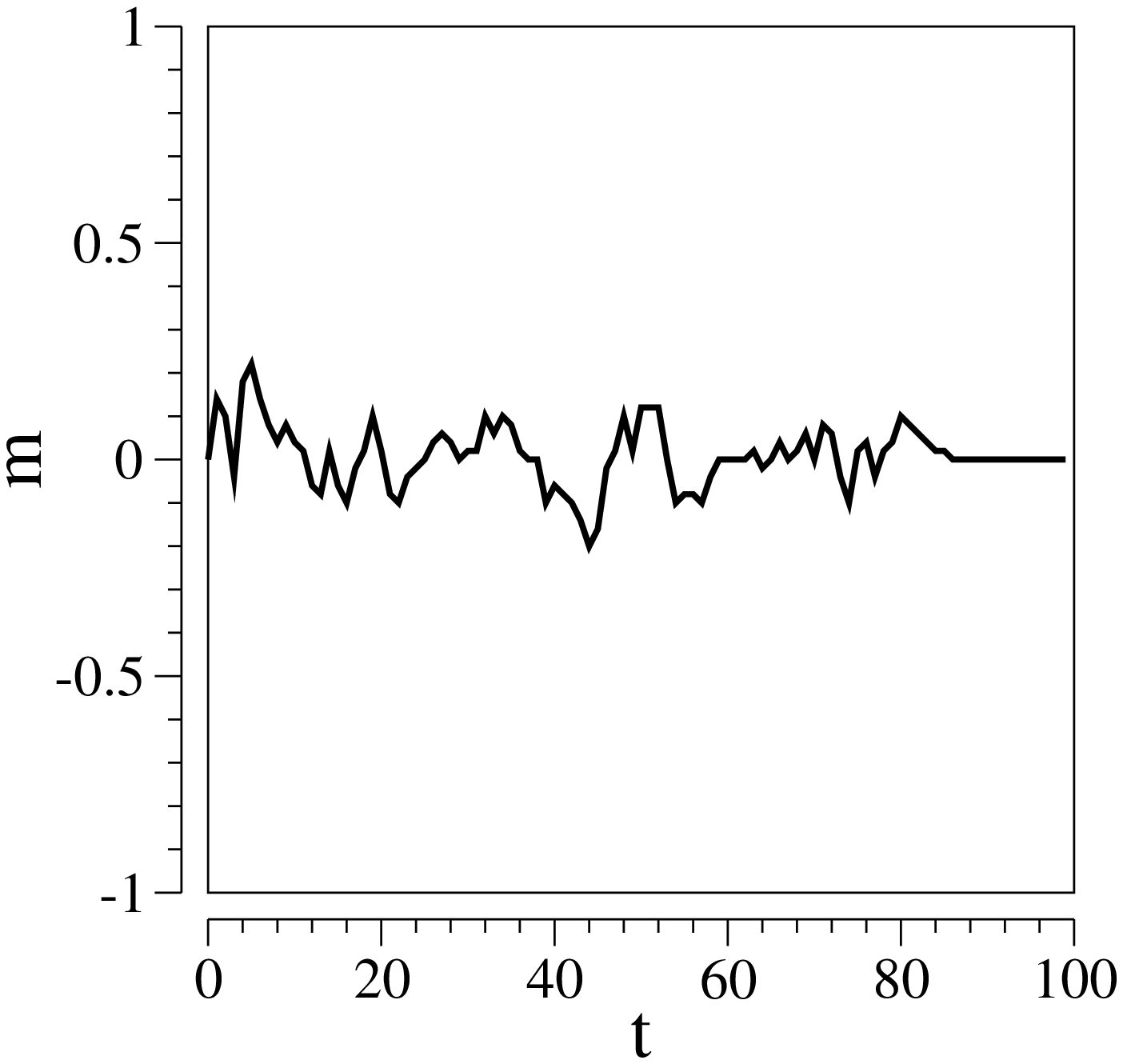,width=8cm}
 }
\vspace*{-10pt}
  \caption{Evolution of the one-dimensional lattice (left) and 
    magnetization $m$ (right) vs time in Monte-Carlo steps.
    Asymptotically, the \emph{coexistence attractor} (or stalemate,
    \emph{antagonistic} attractor) is reached.  (Same parameters as in
    \pic{fig:Snaid_01}, but different random numbers)}
  \label{fig:Snaid_05}
\end{figure}

\begin{figure}[htbp]
\vspace*{-10pt}
\centerline{
\epsfig{file=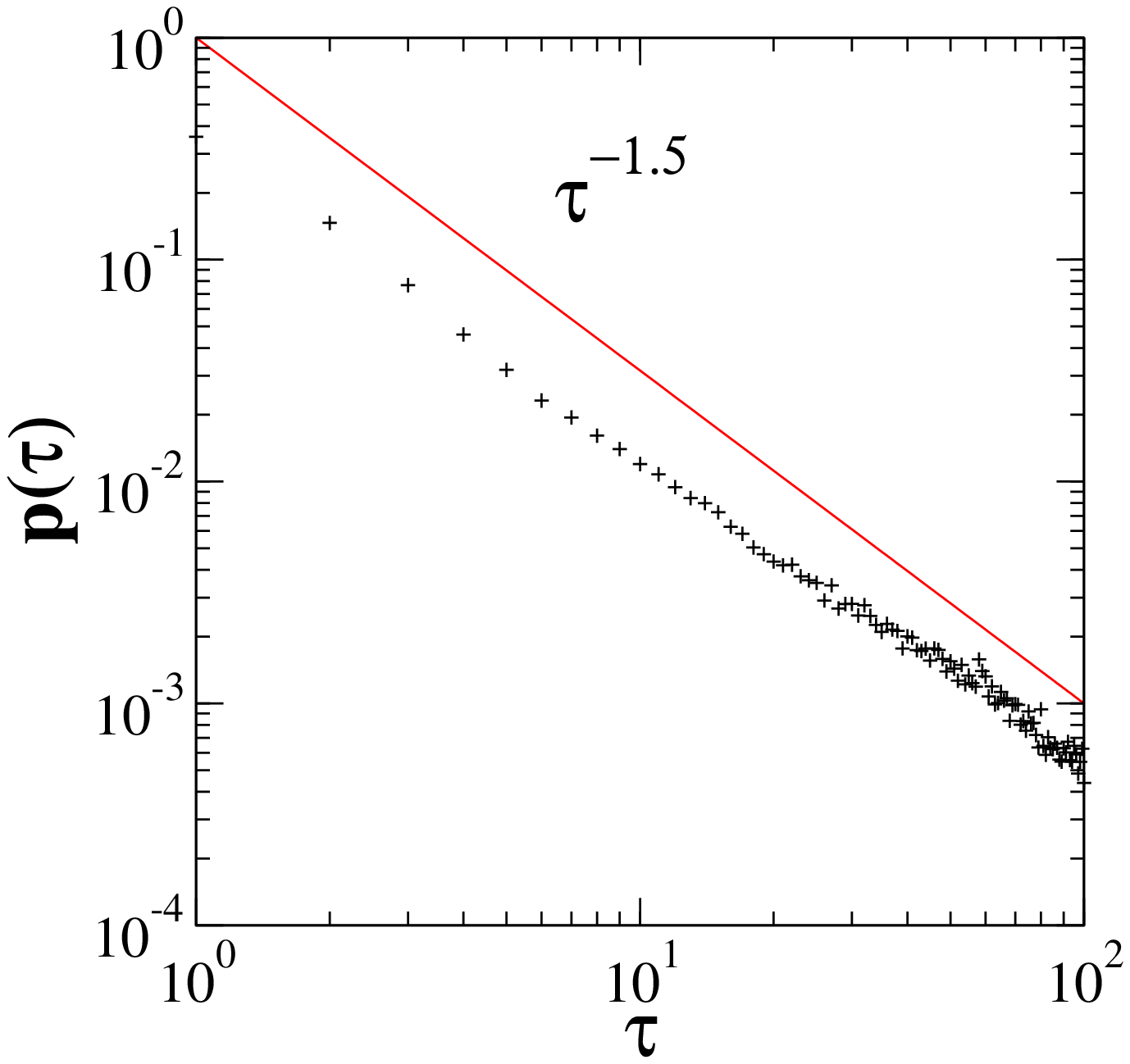,width=8cm}\hfill
\epsfig{file=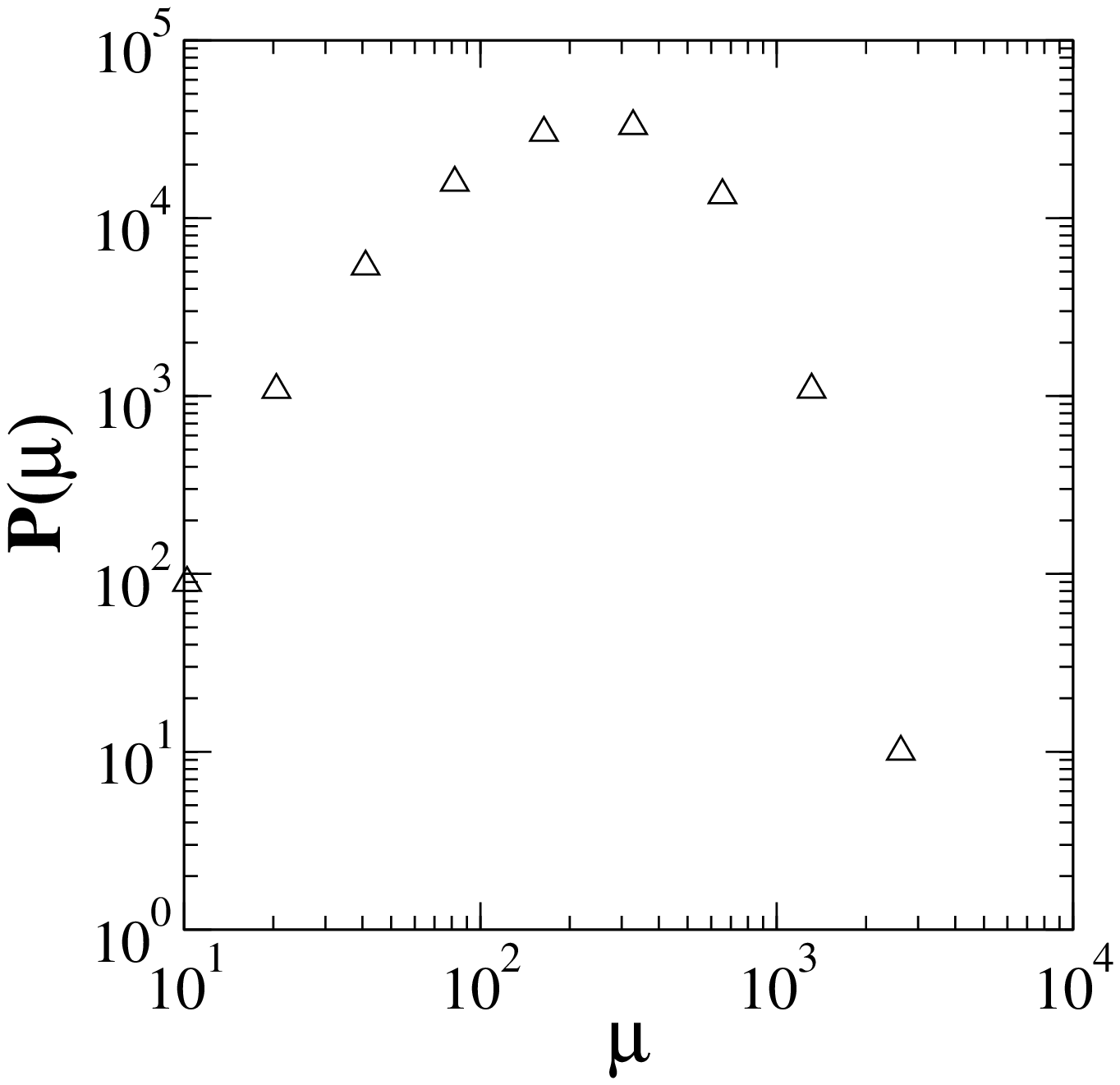,width=8cm}}
\vspace*{-10pt}
  \caption{
    (Left) Distribution of decision times $P(\tau)$ averaged over 200
    simulations, (right) distribution $P(\mu)$ of relaxation times $\mu$
    averaged over 100.000 simulations. ($N=100$, asynchronous update
    according to the Sznajd rules, \eqn{eq:rules}) }
  \label{power-05a}
\end{figure}
In \pic{power-05a}(left) we present the distribution $P(\tau)$ of
decision times $\tau$ introduced in the original paper \citep{weron:2000}
as the time needed by an ``individual'' to change his/her opinion. I.e.,
$\tau$ is a measure of how frequently the opinion of a particular
individual changes, if it is selected by the asynchronous dynamics (which
is apparently not at every flip, but on average once during one
Monte-Carlo step).  The power-law behavior $P(\tau)\propto \tau^{-1.5}$,
already found in \citep{weron:2000} can be clearly observed in
\pic{power-05a}.

Finally, in \pic{power-05a}(right) we also present the distribution
$P(\mu)$ of relaxation times $\mu$ into one of the possible attractors.
The bin size for the histogram has been chosen as $2^{n+1}-2^{n}$, to
allow for comparison with \citep{sousa:2000}. We find that the
distribution has its maximum at about $250$ MC steps, which means that an
average simulation with a CA of $N=100$ needs about this much time. But,
deviations from this mean value follow approximately a \emph{log-normal}
distribution, as shown in \pic{power-05a}(right). This agrees with the
finding of Stauffer for a two-dimensional CA (for rules IIc and III)
\citep{sousa:2000}.

\section{Reformulation of the SM in terms of a VM}
\label{3}
\subsection{Rules of the VM Game}
\label{sec:3.1}

Generally, it is argued that the SM is different from the voter model
(VM) in that the opinion spreads outwards instead of inwards. In this
section, we reformulate the SM in terms of a VM based on the Ising spin
concept and will demonstrate that the SM is in fact a \emph{linear VM}.
This derivation will occur in three steps.

First, we wish to point out the basic idea of a VM. There, the adoption
of an opinion $+1$ or $-1$ of a given site $i$ depends on the \emph{local
  frequency} of the respective opinions in the immediate neighborhood.
Usually, only nearest neighbors are taken into account. The transition
rate for changing $\theta_{i}$ is generally given as:
\begin{equation}
  \label{trans}
  w(\theta_{i}^{\prime}| \theta_{i})=\kappa(f)\,f_{\theta_{i}^{\prime}}
\end{equation}
where $f_{\theta_{i}^{\prime}}$ is the local frequency of the
\emph{opposite} opinion in the nearest neighborhood of site $i$
possessing opinion $\theta_{i}$, and $\kappa(f)$ is a non-linear function
dependent on local frequency. In the linear VM,
$\kappa(f)=\mathrm{const.}$ is chosen.

In order to derive a similar transition rate for the SM, we look at the
possible local configurations in the neighborhood of site $i$, given in
the first column of  \eqn{voter-full}. Here, we have to note a basic
difference between SM and VM. In SM, a pair of sites $\{i-2,i-1\}$
influences its two neighbors $\{i-3,i\}$ at the same time, i.e. the
dynamics of $i$ is influenced only by \emph{one} pair of neigbors
(\emph{either} from the left \emph{or} from the right side). As opposed
to that, in the VM, the local frequency of opinions \emph{both} from the
left \emph{and} from the right side is taken into account. So, looking at
the second nearest neighborhood on both sides, we have a total of 16
possible configurations. The respective transition rates
$w(\theta_{i}^{\prime}| \theta_{i})$ for adjusting $\theta_{i}$ are then
defined in such a way that they lead to the \emph{same dynamics} as in
the SM.

\begin{equation}
  \label{voter-full}
\begin{array}{ccccccc}
\theta_{i-2}& \theta_{i-1} &\theta_{i}& \theta_{i+1}& \theta_{i+2}& 
\quad w(+| \theta_{i}) \quad & w(- | \theta_{i}) \\ \hline
+&+&?& +&+& 1 & 0 \\
-&-&?& -&-& 0 & 1 \\
+&+&?&-&-&0.5 & 0.5 \\
-&-&?&+&+&0.5 & 0.5 \\\hline
-&+&?&+&-&0 & 1 \\
+&-&?&-&+&1 & 0 \\
-&+&?&-&+&0.5 & 0.5 \\
+&-&?&+&-&0.5 & 0.5 \\ \hline
+&+&?&-&+& 1 & 0\\
-&-&?&+&-&0 & 1 \\
+&-&?&+&+& 1 & 0 \\
-&+&?&-&-&0 & 1 \\ \hline
+&+&?&+&-&\sigma & 1-\sigma\\ 
-&+&?&+&+&\sigma & 1-\sigma \\
+&-&?&-&-&1-\sigma & \sigma\\
-&-&?&-&+&1-\sigma & \sigma\\
\end{array}
\end{equation}
In \eqn{voter-full}, the first four transition rates are based on
the ferromagnetic principle, since $\theta_{i-2} \theta_{i-1}=
\theta_{i+1} \theta_{i+2}=1$ (rule 1 of SM).  The next four transition
rates are based on the anti-ferromagnetic principle, since $\theta_{i-2}
\theta_{i-1}= \theta_{i+1} \theta_{i+2}=-1$ (rule 2 of SM). The last
eight possible configurations do not correspond to the SM, because
opinion $\theta_i$ has to choose between ferromagnetism and
antiferromagnetism since $\theta_{i-2}\theta_{i-1}=1$ and $ \theta_{i+1}
\theta_{i+2}=-1$ and vice versa. In particular, in the last four cases
\textrm{frustration} occurs because $\theta_{i}$ cannot simultaneously
accomodate the opinions of both neighboring pairs. For example, if the
pair $++$ appears on the left and the pair $+-$ appears on the right
side, then the opinion bias from the left side would push $\theta_{i}$
towards $+$, while from the right side, it would push $\theta_{i}$
towards $-$.  For those cases, we have introduced a parameter $\sigma$ to
bias the decision towards either the anti- or the ferromagnetic case.
I.e., if $\sigma=0$, opinion $\theta_{i}$ is completely biased by the
anti-ferromagnetic neighbor pair, while for $\sigma=1$ it is completely
biased by the ferromagnetic neighbor pair.  However, if $\sigma=0.5$,
opinion $\theta_i$ is equally balanced between ferromagnetism and
antiferromagnetism spread.

\Eqn{voter-full} can be used as a lookup table for the microsimulations.
But, in order to derive a generalized voter rule, we want to find a
frequency dependent form for the transition rates in the form of
\eqn{trans}. This will be done in the second step, as follows.  The local
frequencies of the different opinions in the \emph{nearest} neighborhood
-- $f^{(1)}_{+}$ -- and in the \emph{second nearest} neighborhood --
$f^{(2)}_{+}$ -- are defined as:
\begin{equation}
  \label{freq}
f^{(1)}_{+}=\frac{1}{2}(\delta_{+;\theta_{i-1}} +
\delta_{+;\theta_{i+1}})\;;\quad 
f^{(2)}_{+}=\frac{1}{2}(\delta_{+;\theta_{i-2}} + \delta_{+;\theta_{i+2}})
\end{equation}
Using the local frequencies, we can reduce the lookup table to only nine
different transition rates, as follows: 
\begin{equation}
  \label{table2}
  \begin{array}{ccccc}
\mbox{}\quad &\quad f^{(1)}_{+} \quad & f^{(2)}_{+} & 
\quad w(+| \theta_{i}) \quad & w(- | \theta_{i}) \\ \hline
1. & 1 & 1 &1&0\\
2. & 0&0&0&1\\ \hline
3. & 0.5 & 0.5 &0.5&0.5\\ \hline
4. & 1 &0&0&1\\ 
5. & 0&1&1&0\\ \hline 
6. & 0.5& 0 &0&1 \\
7. & 0.5 & 1&1&0\\ \hline
8. & 1 & 0.5 &\sigma&1-\sigma\\
9. & 0& 0.5 &1-\sigma&\sigma\\
\end{array}
\end{equation}
The first three transition rates (1.-3.) apply in cases where
$\theta_{i}$ has ferromagnetic pairs on \emph{both} sides, while the
three cases (3.-5.) apply if $\theta_{i}$ has anti-ferromagnetic pairs on
\emph{both} sides. The last four cases (6.-9.) apply if $\theta_{i}$ has
one anti- and one ferromagnetic pair on each side. Again, cases 8. and 9.
are special in the sense that \emph{frustration} occurs because of the
discrepancy between the opinion biases from the left and from the right
side.

In the third step, we conclude the frequency dependence of the transition
rates given in \eqn{table2} in a most concise equation: 
\begin{eqnarray}
\label{eq:nvoter}
w(+|\theta_i)=  \kappa(f_{+})\,f^{(2)}_{+}&; \;& 
\kappa(f_{+}) =
\left \{
\begin{array}{cc}
1 & \quad \mbox{(no frustration)} \\ 
2 \sigma \,f^{(1)}_{+} + 2(1-\sigma)\, (1- f^{(1)}_{+})
& \quad \mbox{(frustration)} 
\end{array}
\right. \nonumber \\ 
&&  \\ 
w(-|\theta_i)=  \kappa(f_{-})\,f^{(2)}_{-}&; \; &
\kappa(f_{-}) =
\left \{
\begin{array}{cc}
1 & \quad \mbox{(no frustration)} \\ 
2 \sigma \,f^{(1)}_{-} + 2(1-\sigma)\, (1- f^{(1)}_{-})
& \quad \mbox{(frustration)} 
\end{array}
\right. \nonumber
\end{eqnarray}
It turns out that the dynamics of the SM can be rewritten in terms of a
VM where the frequency dependent transition rate basically depends on the
frequencies of opinions of the \emph{second nearest neighbors},
$f^{(2)}_{+}$, $f^{(2)}_{-}$. The opinion frequencies of the first
nearest neighbors, $f^{(1)}_{+}$, $f^{(1)}_{-}$ only enter the prefactor
$\kappa$, thus making the transition rate a non-linear voter rule.

We note that $\kappa$ is different from 1 only in the case of
\emph{frustration}, in which the nearest neighbor frequencies
$f^{(1)}_{+}$, $f^{(1)}_{-}$ can have only values of 0 or 1.  However,
for the special case of $\sigma=0.5$, i.e. no bias towards either anti-
or ferrromagnetism, $\kappa$ becomes 1 even in the case of frustration.
Thus, we can reduce the dynamics of \eqn{eq:nvoter} to a \emph{linear}
voter rule, valid for all cases:
\begin{equation}
\label{eq:voter2}
w(+|\theta_i)=  f^{(2)}_{+}\;; \quad 
w(-|\theta_i)=  f^{(2)}_{-} \quad(\mathrm{for}\; \sigma=0.5)
\end{equation}
This remarkable finding is based on theoretical investigations of the
possible local configurations -- but in \sect{3.2} we will show by means
of computer simulations that the linear rule of \eqn{eq:voter2} matches
numerically with the dynamics of the SM.

To summarize, the only VM \emph{transition rates that matter for the
  simulation of the SM}, are simply given by the three cases:
\begin{equation}
  \label{table3}
  \begin{array}{ccc}
f^{(2)}_{+}  & \quad w(+| \theta_{i}) \quad & w(- | \theta_{i}) \\ \hline
1&1&0\\
0 & 0 &1\\
0.5 & 0.5 &0.5\\
\end{array}
\end{equation}
The additional transition rates given in \eqn{table2} depend on
$\sigma$ and result from the possibility of considering frustration
dynamics, which exceeds the original idea of the SM. 

Finally, we emphasize that the importance of the second neighbors on the
opinion dynamics is already a basic ingredient of the SM (even if not
seen that way).  Thus, the two rules of the SM can be simply combined
into only \emph{one} rule, namely \emph{``just follow your second nearest
  neighbor''}. Specifically, using the notation of the SM,
$\theta_{i-1}=\theta_{i+1}$ and $\theta_{i+2}=\theta_i$, no matter
whether $\theta_i \theta_{i+1}=1$ or $\theta_i \theta_{i+1}=-1$.  To
repeat this important finding, in SM the nearest neighbors of a site are
just ignored in the dynamics.

\subsection{Results of VM Computer Simulations}
\label{3.2}

In this section, we compare the results of the VM dynamics with the known
results of the SM. Therefore, we fix $\sigma=0.5$ because only in this
case the VM is equivalent to the SM, according to the previous section.
The figures showing the spatio-temporal evolution of the lattice states
and the respective magnetization shall be compared with the corresponding
figures obtained from the SM. The basic setup chosen is the same, i.e.
$N=100$ and periodic boundary conditions for the lattice, initially
uniform random distribution of the opinions, and asynchronous update
rule.

As we can see in \pic{fig:votera_05} and \pic{fig:votera_01}, the
dynamics eventually reach either the coexistence or the consensus
attractor, and even the intermediate coexistence of anti- and
ferromagnetic domains can be observed as in the case of SM. Also, the
power law of the distribution of decision times $P(\tau)$,
\pic{power_voter}(left), remains the same, as well as the distribution of
relaxation times $P(\mu)$, \pic{power_voter}(right), which follows the
log-normal distribution. The only difference to be noticed is that the
average relaxation time has now doubled in VM, compared to SM. This is to
be expected since each update makes two flips in SM, while it makes only
one flip in VM.

\begin{figure}[htbp]
\vspace*{-10pt}
\centerline{\epsfig{file=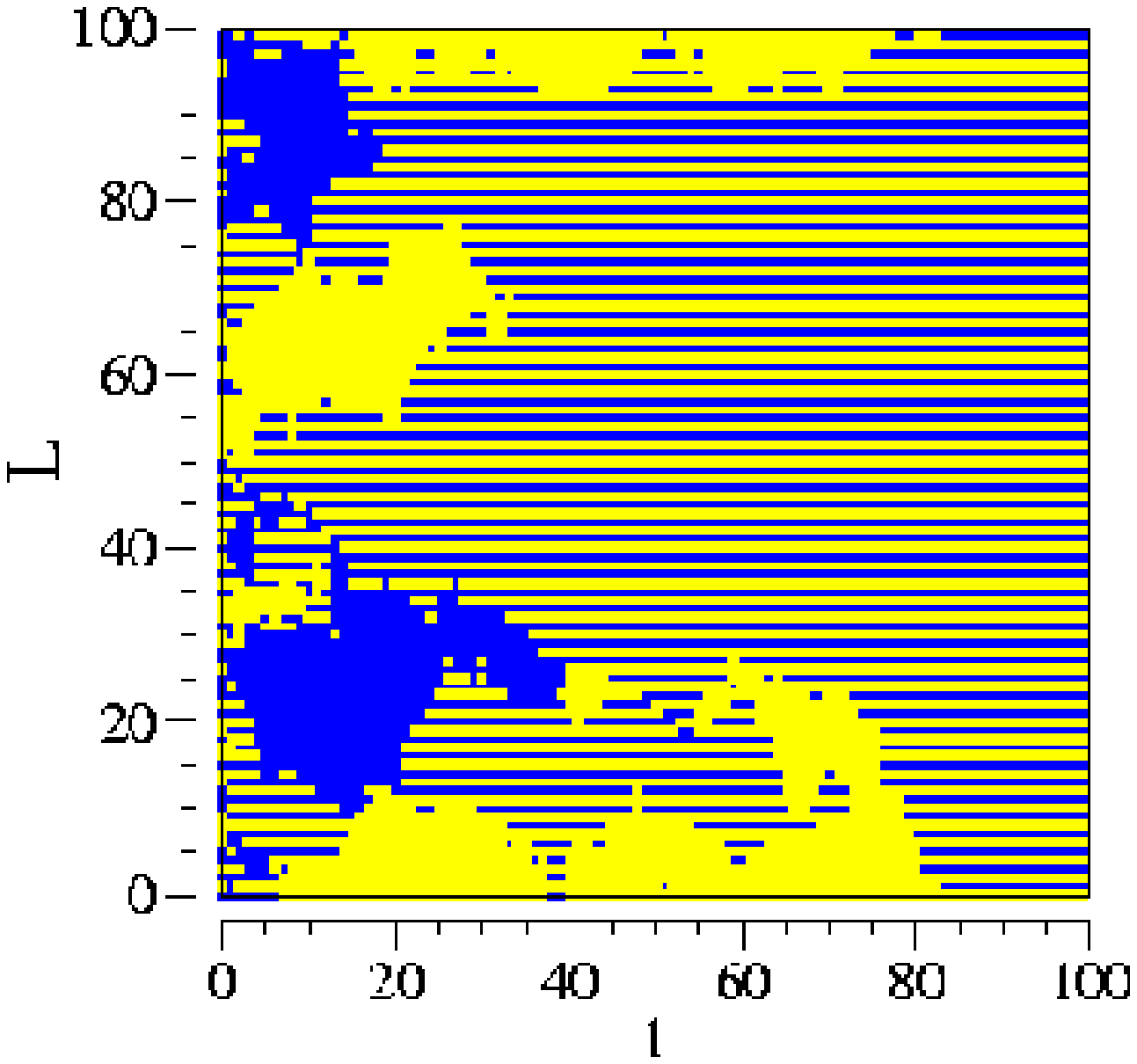,width=8cm} \hfill
\epsfig{file=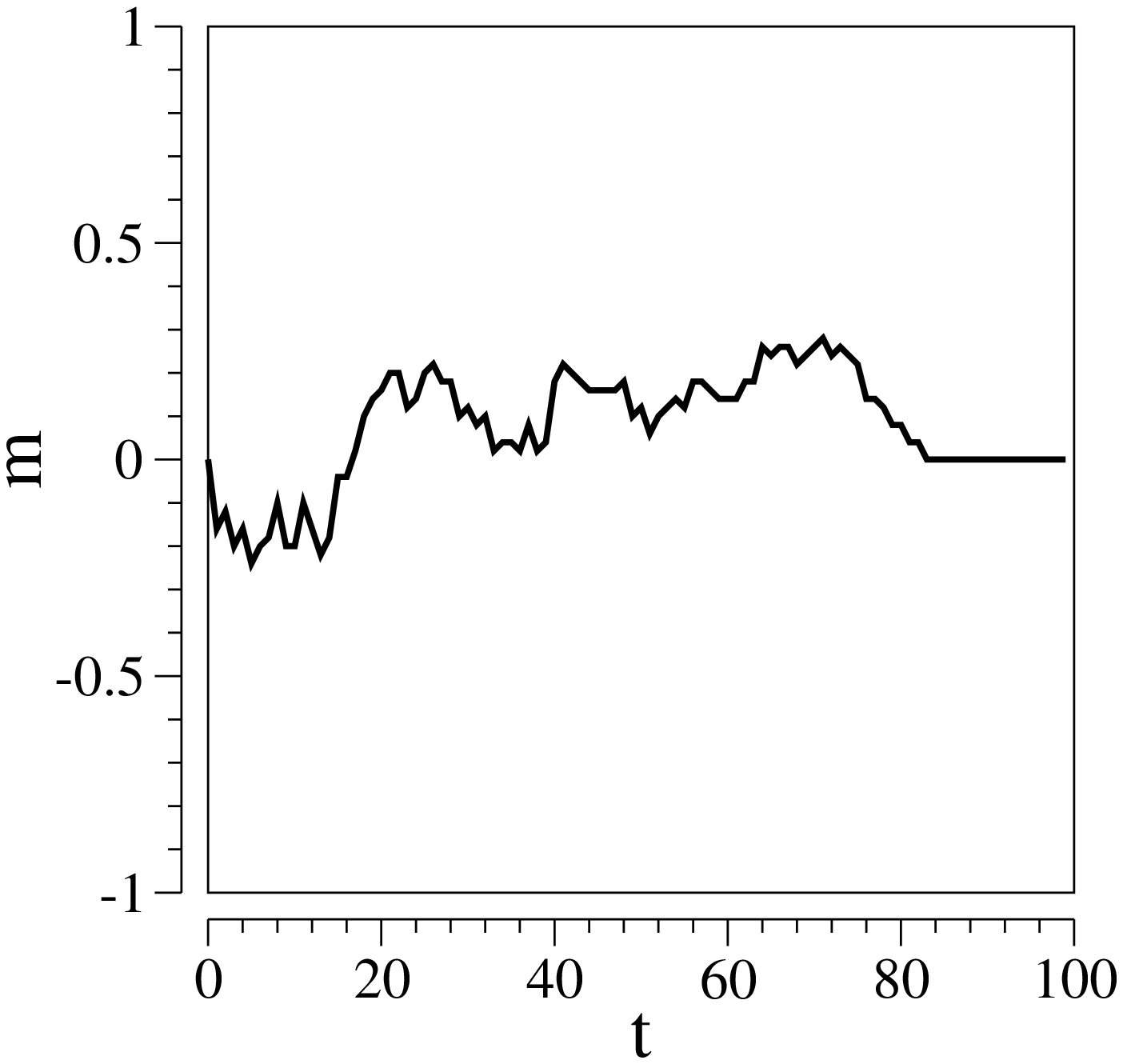,width=8cm}
 }
\vspace*{-10pt}
  \caption{Evolution of the one-dimensional lattice (left) and 
    magnetization $m$ (right) vs time in Monte-Carlo steps.
    Asymptotically, the \emph{coexistence attractor} is reached.
    (Same setup as in \pic{fig:Snaid_05}, but dynamics according to the
    linear voter rule, \eqn{eq:voter2}.)  }
  \label{fig:votera_05}
\end{figure}

\begin{figure}[htbp]
\vspace*{-10pt}
\centerline{
\epsfig{file=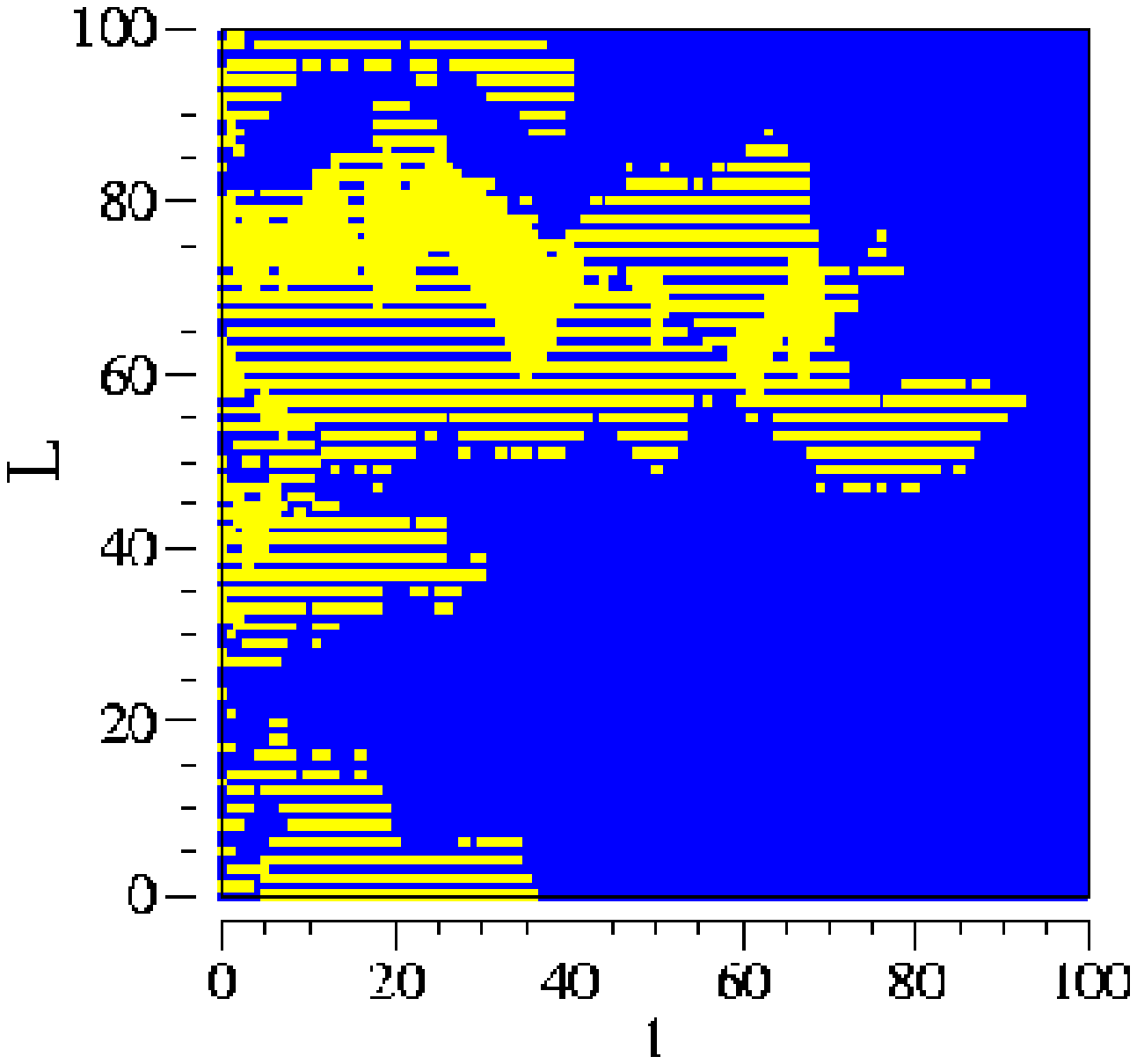,width=8cm}\hfill 
\epsfig{file=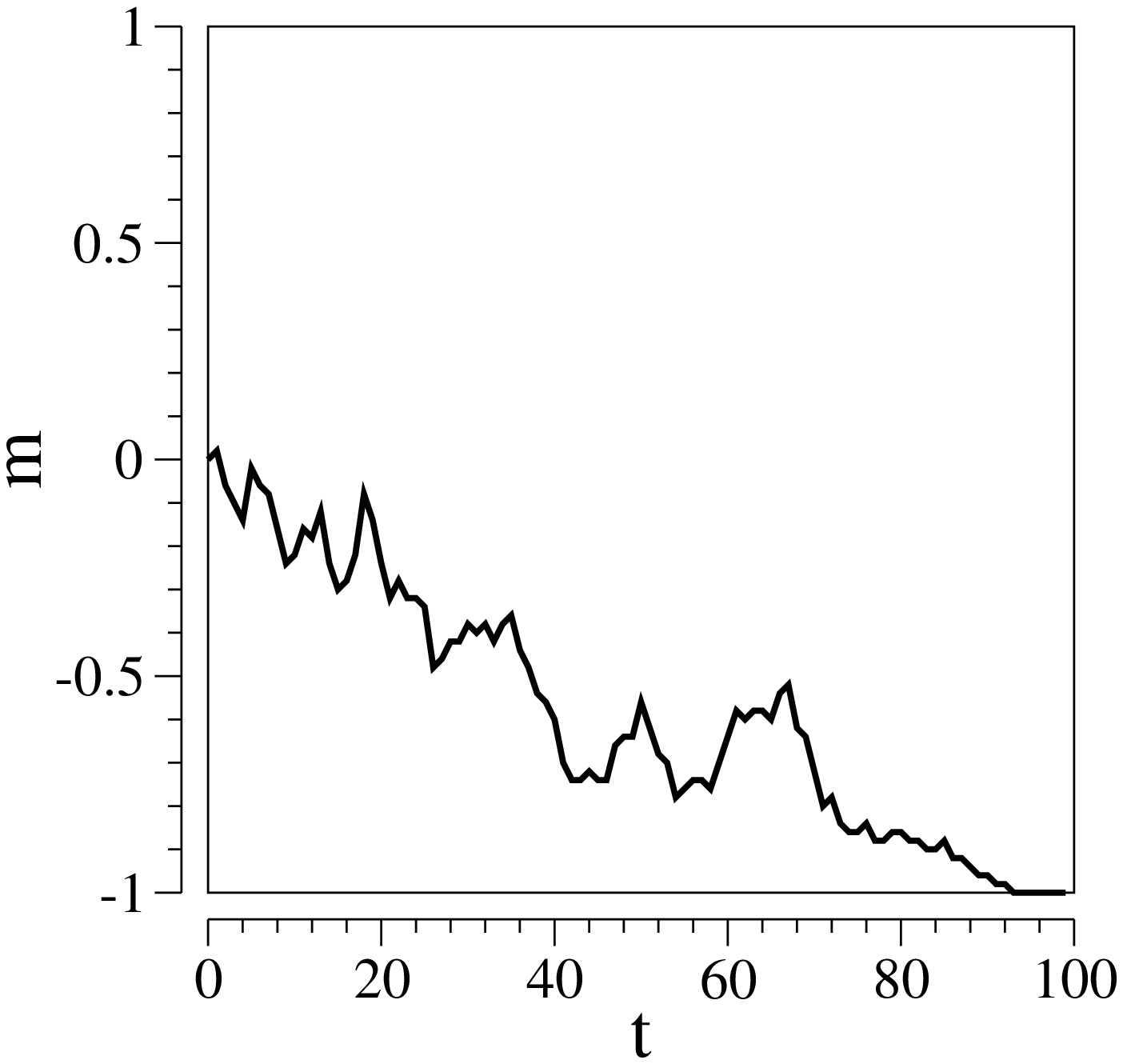,width=8cm} 
}
\vspace*{-10pt}
  \caption{Evolution of the one-dimensional lattice (left) and 
    magnetization $m$ (right) vs time in Monte-Carlo steps.
    Asymptotically, the \emph{concensus attractor} is reached. (Same
    setup as in \pic{fig:Snaid_01}, but dynamics according to the linear
    voter rule, \eqn{eq:voter2}.) }
  \label{fig:votera_01}
\end{figure}

\begin{figure}[htbp]
\vspace*{-10pt}
\centerline{
\epsfig{file=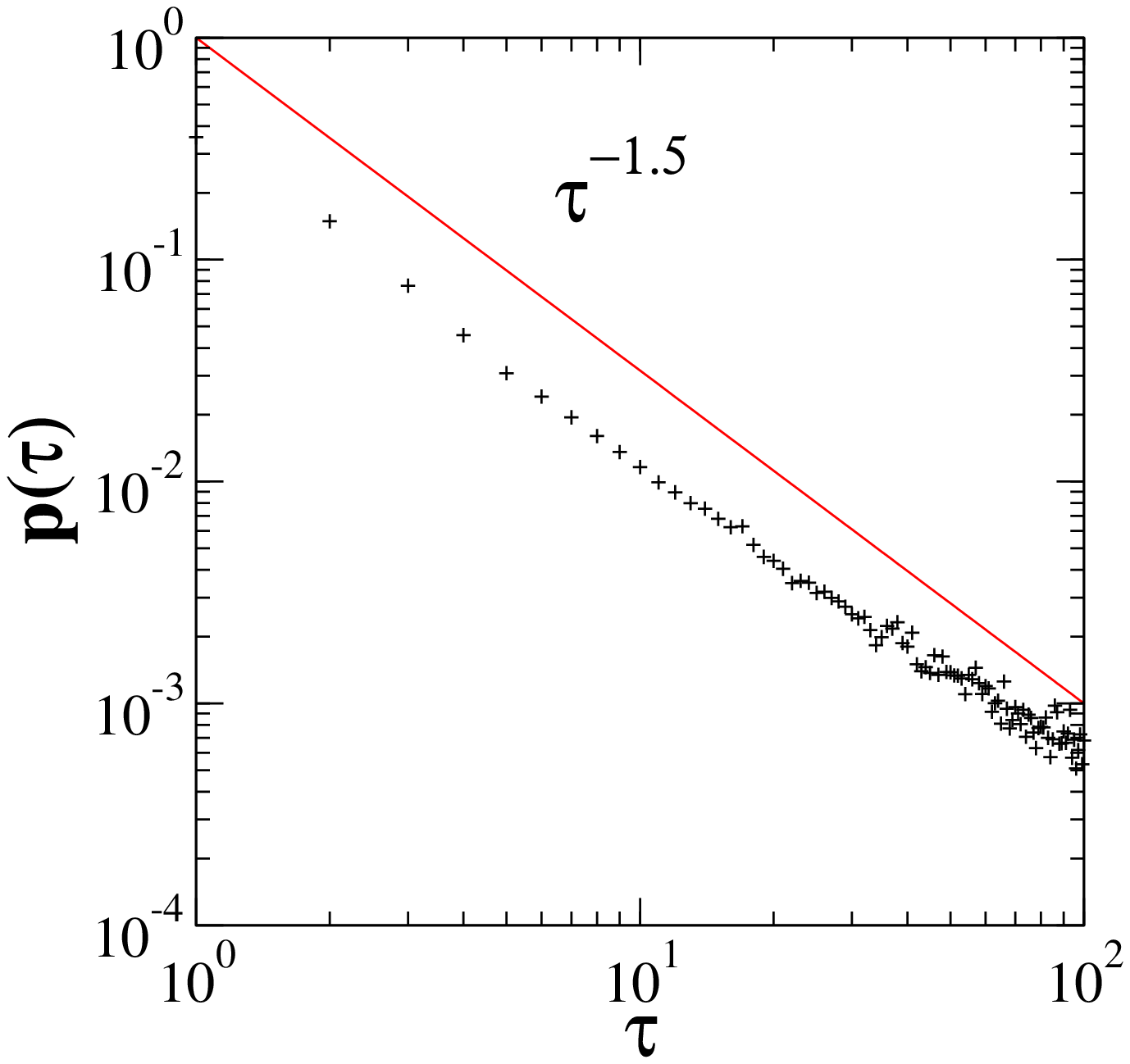,width=8cm}\hfill
\epsfig{file=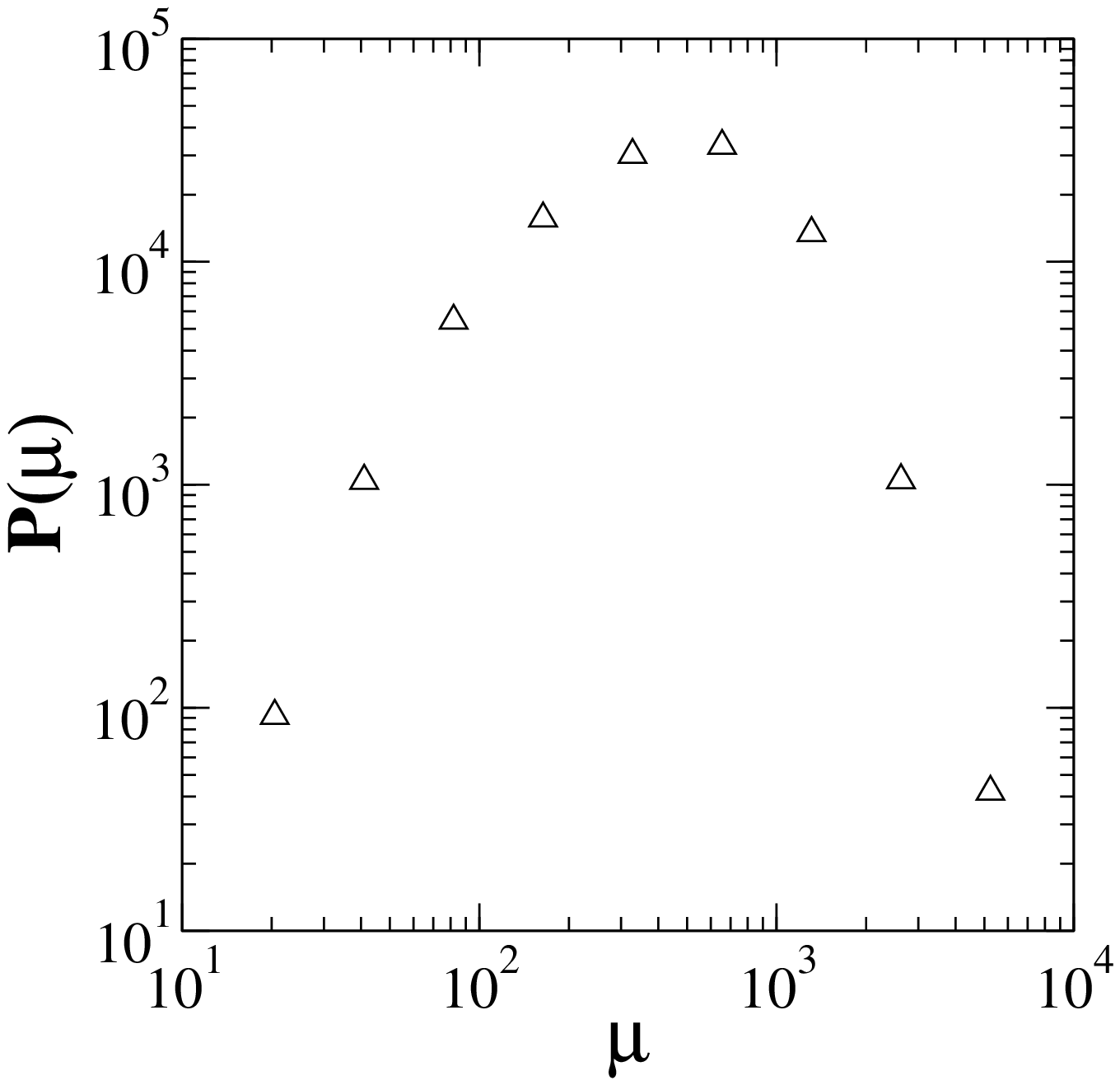,width=8cm}}
\vspace*{-10pt}
  \caption{
    (Left) Distribution of decision times $P(\tau)$ averaged over 200
    simulations, (right) distribution $P(\mu)$ of relaxation times $\mu$
    averaged over 100.000 simulations (dynamics according to the linear
    voter rule, \eqn{eq:voter2}.)}
  \label{power_voter}
\end{figure}

Thus, our microsimulations show that the proposed linear voter rule
($\sigma=0.5$) is not different from the SM, both in terms of the
dynamics and the final attractors.  In the next section, we will show
that this holds also for the frequency of reaching the attractors (50
percent stalemate, 25 percent up and 25 percent down).

We conclude that it does not matter, whether ``the influence flows inward
from the surrounding neighbors to the center site, or spreads outwards in
the opposite direction from the center to the neighbors'' as argued in
\citep{sousa:2000,elgazzar:2001}, i.e. there are basically no principle
differences between the SM and the VM except in the expression of the
rules.

\section{Influence of the Bias Parameter $\sigma$}
\label{3.3}

When comparing SM and VM, we found that the decision dynamics for some
local configurations is characterized by some sort of \emph{frustration},
because of a conflict between the left and right opinion bias.  In order
to break the symmetry in those cases, we have introduced the bias
parameter $\sigma$, which favors the anti-ferromagnetic response for
$\sigma\to 0$ and the ferromagnetic response for $\sigma \to 1$. Only for
the case of $\sigma=0.5$, no bias is given -- which is  the 
case for expressing the SM in terms of a VM.

In this section, we want to pay more attention to the role of $\sigma$,
which exceeds the original idea of the SM. Let us first look at the
probability of reaching the different attractors. We recall from \sect{2}
that in SM three attractors exist, where the two ferromagnetic attractors
are reached with probability 0.25 each, whereas the anti-ferromagnetic
attractor is reached with probability 0.5. In \sect{3.2}, we have already
shown that these attractors are also reached in the case of a linear
voter model ($\sigma=0.5$). To estimate the probability, we run 1.000
computer simulations with different values of $\sigma$ (Table \ref{1000})

\begin{table}[htbp]
  \begin{center}
\begin{tabular}{cccccccc}
$\sigma$ & 0.00 & 0.25 & 0.45 & 0.50 & 0.55 & 0.75 & 1.00 \\ \hline 
$f_{-+}$ & 1 & 1 & 0.998 & 0.510  & 0.004 & 0   & 0    \\ \hline
$f_{++}$ & 0 & 0 & 0.002     & 0.250 & 0.526 & 0.496 & 0.510     \\
$f_{--}$ & 0 & 0 & 0     &  0.240    &  0.470    & 0.504     & 0.490     \\ \hline
    \end{tabular}
  \end{center}
    \caption{
      Frequency of reaching the different attractors,
      \emph{ferro}$_{+}$ ($f_{++}$), \emph{ferro}$_{-}$ ($f_{--}$), and
      \emph{anti-ferro} ($f_{-+}$) obtained from 1.000
      simulations ($N=100$, voter rules of \eqn{eq:nvoter},
      asynchronous update).}
    \label{1000}
\end{table}

For $\sigma=0.5$, we observe that in the VM case the three attractors are
reached with the same probability as in the SM case. Thus we can conclude
that there are no differences between SM and VM with respect to this
feature either.

Since $\sigma\to 0$ biases the dynamics towards the anti-ferromagnetic
attractor while $\sigma\to 1$ biases towards the ferromagnetic one, the
VM provides a simple possibility to avoid either stalemate or consensus
in decision making. It is interesting to note that already \emph{small
  deviations} from $\sigma=0.5$ will lead to drastical changes in the
probabilities of reaching the different attractors. I.e., already for
$\sigma=0.45$ or $\sigma=0.55$ only one attractor is found (where the
ferromagnetic one appears in two different ``flavors'').

The disappearance of one of the attractors basically results from a
competition process between anti- and ferromagnetic domains. If a site is
selected \emph{within} a domain, nothing changes. The important events
for the spatio-temporal evolution occur only at the borders
\emph{between} these domains, i.e.  $\{++++{\,?\,}+-+-\}$ or
$\{----{\,?\,}-+-+\}$, where frustration also occurs.  Dependent on the
value of $\sigma$, the following possibilities for the dynamics exist:
\begin{equation}
  \label{update}
++++ {\,?\,} +-+-  \Rightarrow \left \{
\begin{array}{cl}
++++ -+-+- & \quad \mbox{if} \; \sigma=0.0 \\ 
++++ ++-+- & \quad \mbox{if} \; \sigma=1.0 \\ 
++++ {\,?\,} +-+- & \quad \mbox{if} \; \sigma=0.5 \\ 
\end{array}
\right. 
\end{equation}
That means that for $\sigma\to 0$ the anti-ferromagnetic domain will
always increase at the cost of the ferromagnetic one, while for
$\sigma\to 1$ the opposite will occur. Only for $\sigma=0.5$, both cases
occur with the same probability, i.e. in half of the cases the system may
eventually reach (one of) the ferromagnetic attractors and in half of the
cases the anti-ferromagnetic one. We note that this insight, how one
domain may invade the other one, became clear only in the VM picture
through investigation of the frustration dynamics (while it was not
apparent in the SM view). Thus, we conclude that $\sigma$ plays a crucial
role in explaining the \emph{phase transition} known in SM, from the initial
random distribution to either antagonistic or consensus attractor.

\begin{figure}[htbp]
\vspace*{-10pt}
  \centerline{
    \psfig{file=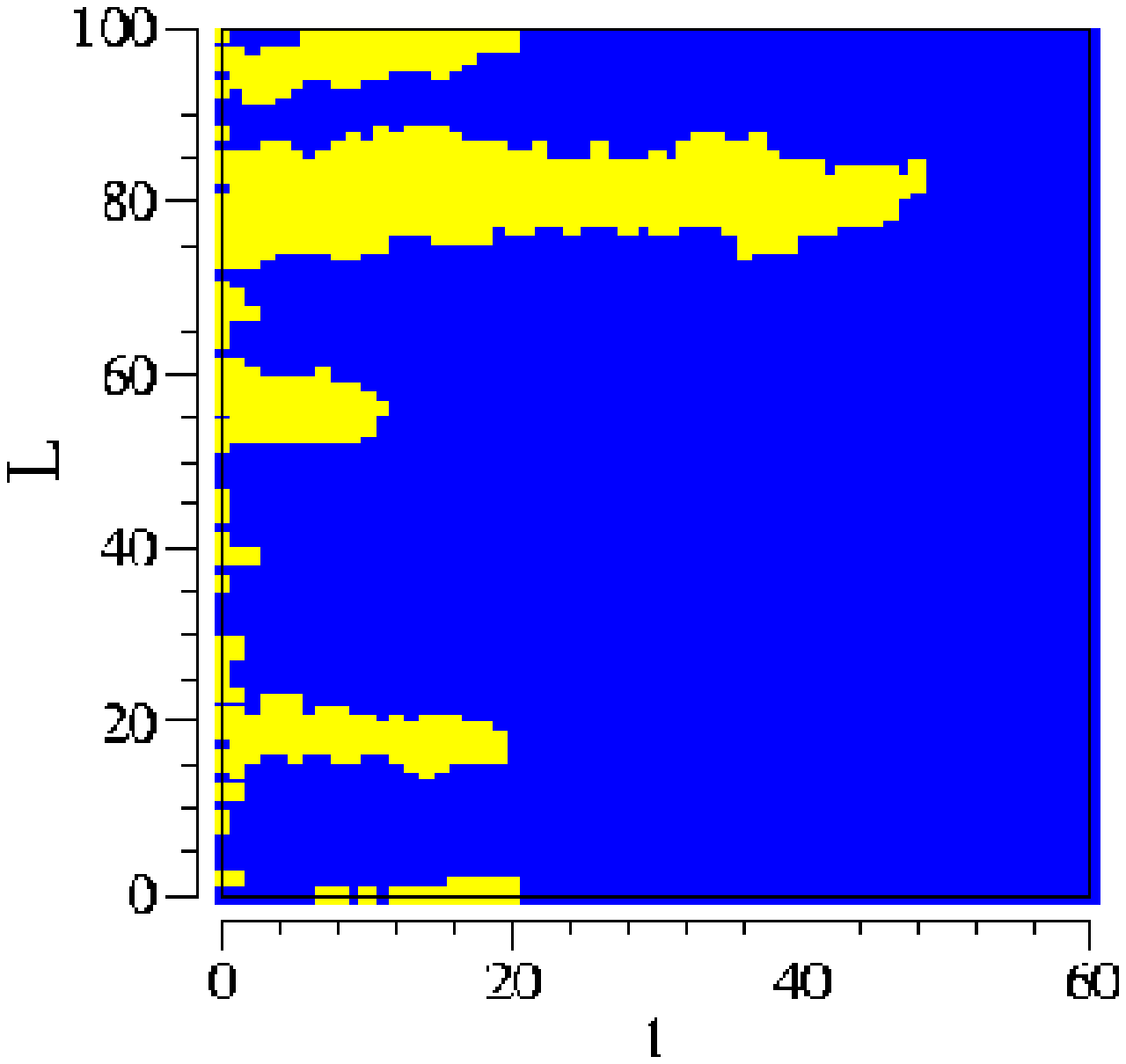,width=8cm} \hfill
    \psfig{file=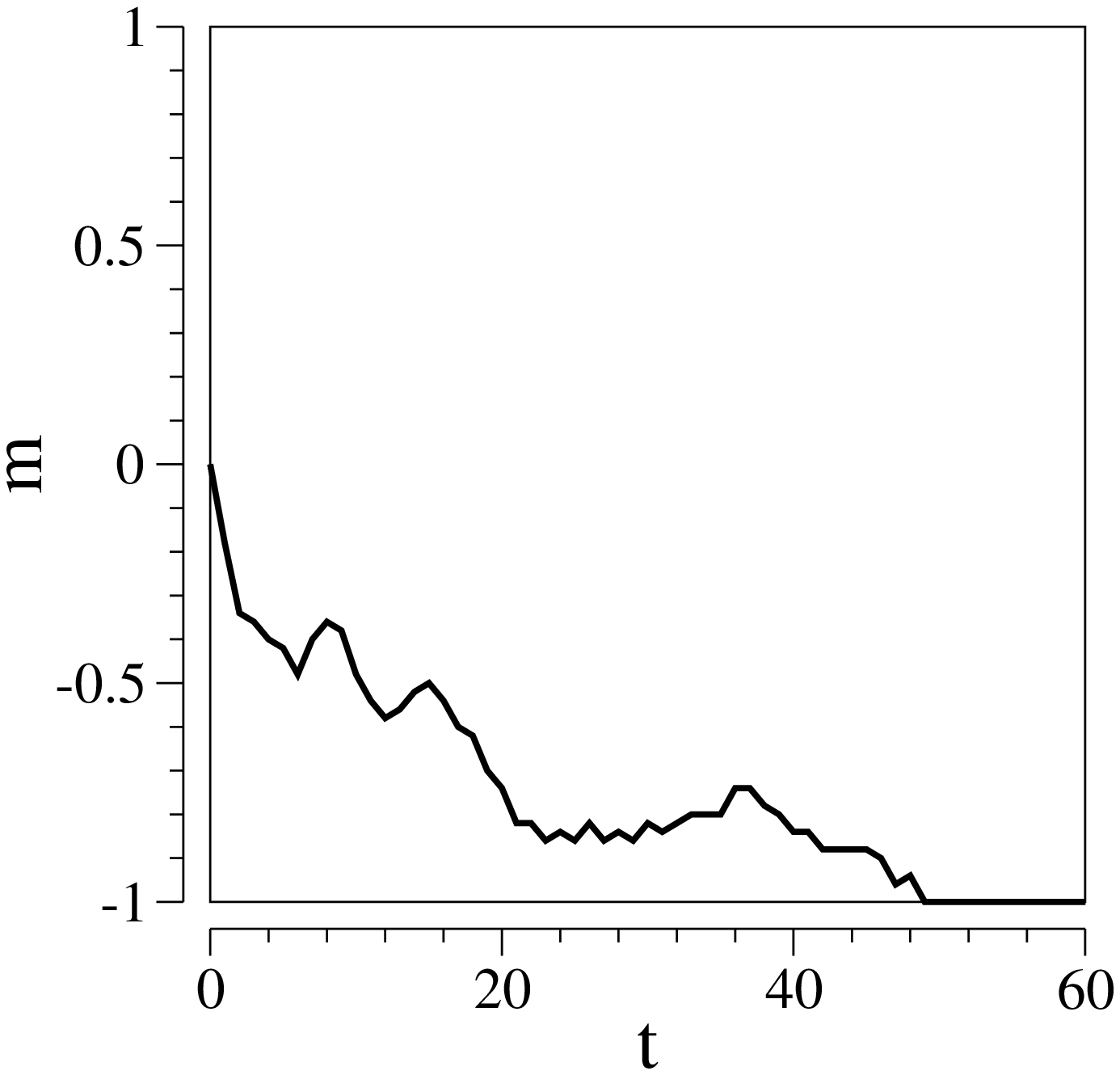,width=8cm}}
\vspace*{-10pt}
  \caption{
    Evolution of the one-dimensional lattice (left) and magnetization $m$
    (right) vs time in Monte-Carlo steps.  Asymptotically, the
    \emph{coexistence attractor} is reached.  (Same setup as in
    \pic{fig:Snaid_01}, but dynamics according to the
    voter rules, \eqn{table2} with $\sigma=1.0$)
  \label{fig:Voter_01}}
\end{figure}

\begin{figure}[htbp]
\vspace*{-10pt}
   \centerline{\psfig{file=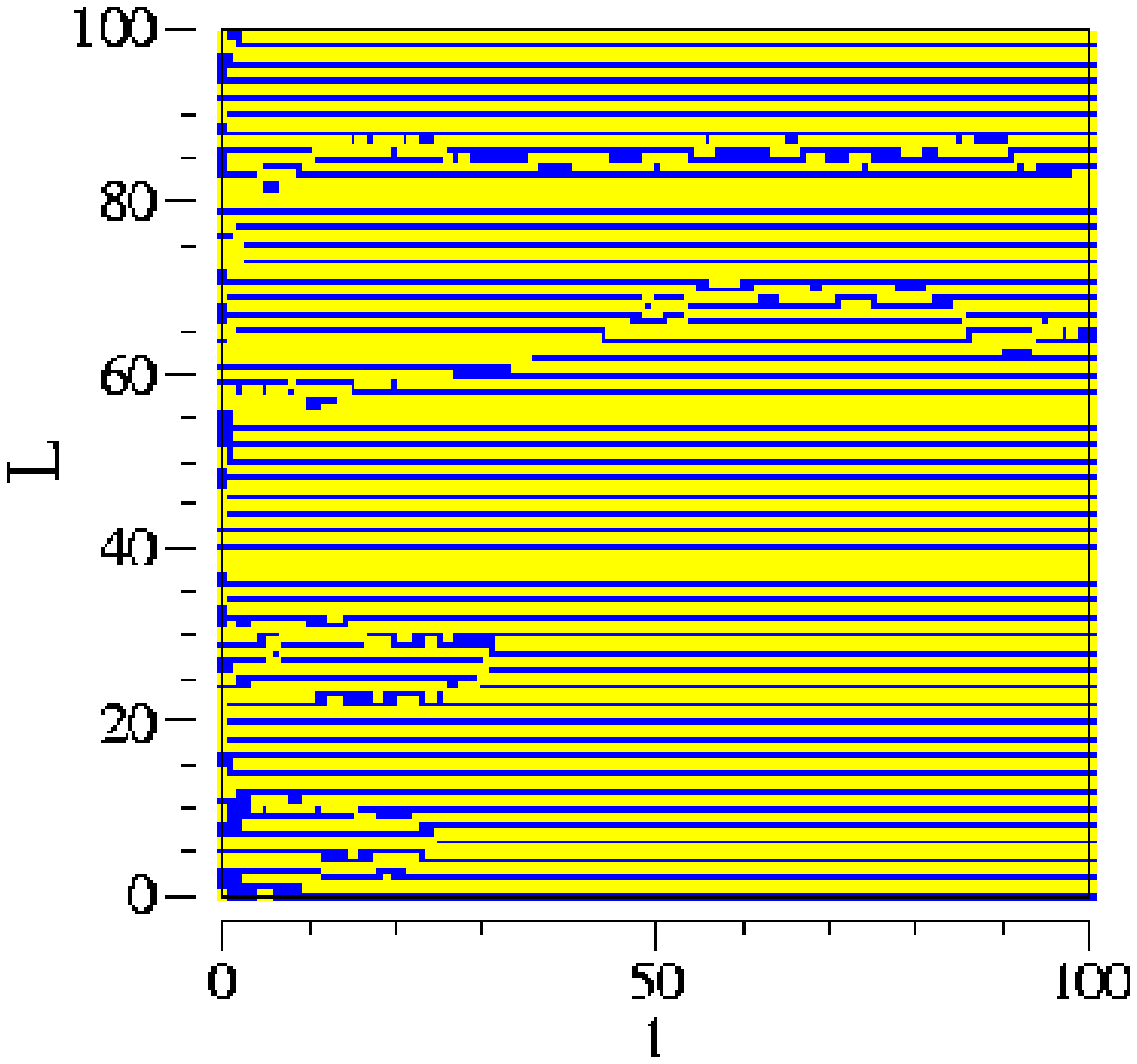,width=8cm}
     \epsfig{file=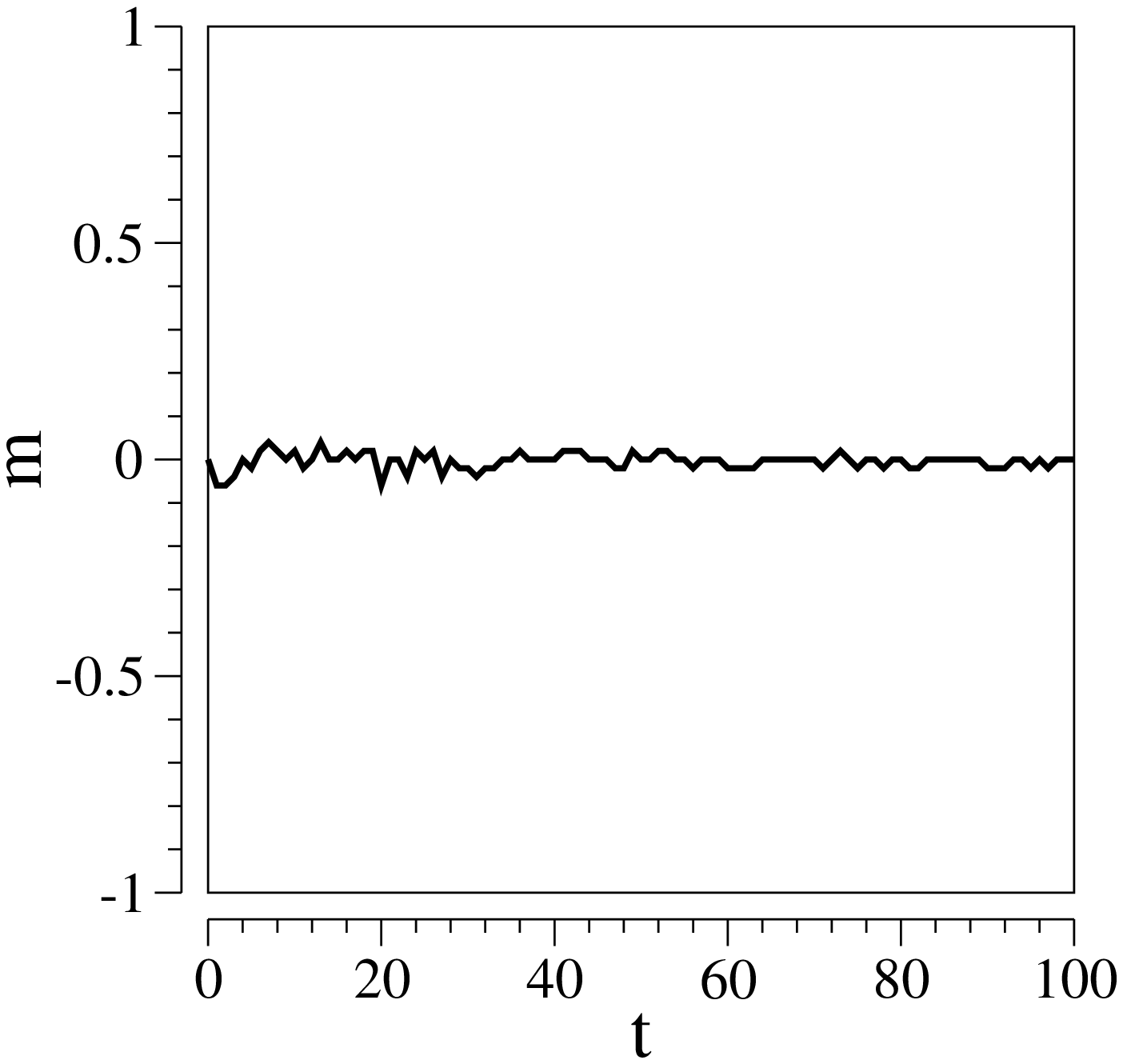,width=8cm} }
\vspace*{-10pt}
   \caption{Evolution of the one-dimensional lattice (left) and 
     magnetization $m$ (right) vs time in Monte-Carlo steps.
     Asymptotically, the \emph{coexistence attractor} is reached.  (Same
     setup as in \pic{fig:Snaid_05}, but dynamics according to the voter
     rules, \eqn{table2} with $\sigma=0.0$)
 \label{fig:Voter_05}}
\end{figure}
 
Finally, we will have a look at the evolution of the spatio-temporal
patterns for the two extreme cases, $\sigma\to 0$, \pic{fig:Voter_01},
and $\sigma \to 1$, \pic{fig:Voter_05}, and compare them with the case of
$\sigma=0.5$, Figs. \ref{fig:votera_05}, \ref{fig:votera_01} where the VM
dynamics are equivalent to the SM dynamics. As already noticed above, the
asymptotic distributions are ``preselected'' by the choice of $\sigma$,
so it is not surprising that either the coexistence or the consensus
attractors are reached.  However, looking at the intermediate dynamics,
we realize that there are no anti-ferromagnetic domains (``striped
patterns'') for $\sigma=1.0$, while there are no ferromagnetic domains
(``filled patterns'') for $\sigma=0.0$. I.e., there is no coexistence
between ferromagnetic and anti-ferromagnetic domains in the intermediate
dynamics in the biased case, while it can be observed in the non-biased
case.

\section{Synchronous vs. Asynchronous Update}
\label{4}

So far, we have used the so-called asynchronous (i.e. random sequential)
update dynamics both for the SM and the VM, which means that at each time
step one lattice site is randomly updated and changes are immediately
processed to the neighborhood. VM, however, were first considered in a
biological context, where time is measured in \emph{generations} and
changes of the lattice states become effective only \emph{after} a
generation is completed (i.e., usually after all sites are selected).
The information generated will be thus processed in \emph{parallel},
which is known as synchronous update.

In this section, we want to investigate whether the different update
rules, i.e. the different ways of information processing, may affect the
outcome of the SM/VM dynamics. Therefore, we have fixed $\sigma=0.5$.
First, we have a look again at the possible attractors of the dynamics,
as shown in Table \ref{6attr}.
\begin{table}[htbp]
  \begin{center}
\begin{tabular}{cccc}
attractor  & \quad local configuration \quad & $f_{+}/f_{-}$ &
\quad frequency \quad \\ \hline
1 & $++++++++++++$ & 1/0 &0.056  \\
2 & $------------$ & 0/1 & 0.060   \\
3 & $-+-+-+-+-+-+$ & 0.5/0.5 & 0.125   \\ \hline
4 & $+++-+++-+++-$ & 0.75/0.25 & 0.250 \\
5 & $---+---+---+$ & 0.25/0.75 & 0.252 \\
6 & $--++--++--++$ & 0.5/0.5 & 0.257 \\ \hline
    \end{tabular}
  \end{center}
    \caption{
      Attractors of the VM dynamics and frequencies of
      reaching them (obtained from 10.000 simulations). $f_{+}/f_{-}$
      gives the frequencies of each opinion in the asymptotic
      configuration. 
      ($N=100$, voter rules of \eqn{eq:nvoter}, $\sigma=0.5$, 
      \textbf{synchronous} update). }
    \label{6attr}
\end{table}

Compared to the asynchronous VM/SM dynamics, we notice the appearance of
three more attractors in the synchronous case. Two of these are different
from the known ones in that they are characterized by a \emph{asymmetric}
coexistence of the two different opinions. I.e., we find stable
configurations (in the absense of noise) with 75 percent of one of the
opinions.  The third new attractor is again a stalemate, or antagonistic
attractor, but the local configuration is different from the known
anti-ferromagnetic one, (\emph{anti-ferro}), attractor 3 in Table
\ref{6attr}.

Furthermore, from the frequencies with which the attractors are reached,
Table \ref{6attr}, we note that the three new attractors are reached with
the same probability of about 0.25, while the ``old'' attractors (1-3)
alltogether only have a probability of 0.25. Again, within that share,
the probability of the anti-ferromagentic attractor is equal to the
probability of both the ferromagnetic ones, 0.125. But, given the
absolute probability, pure anti- (3) and ferromagnetic configurations
(1,2) become rare in synchronous update.

\begin{figure}[htbp]
\vspace*{-10pt}
\centerline{
\epsfig{file=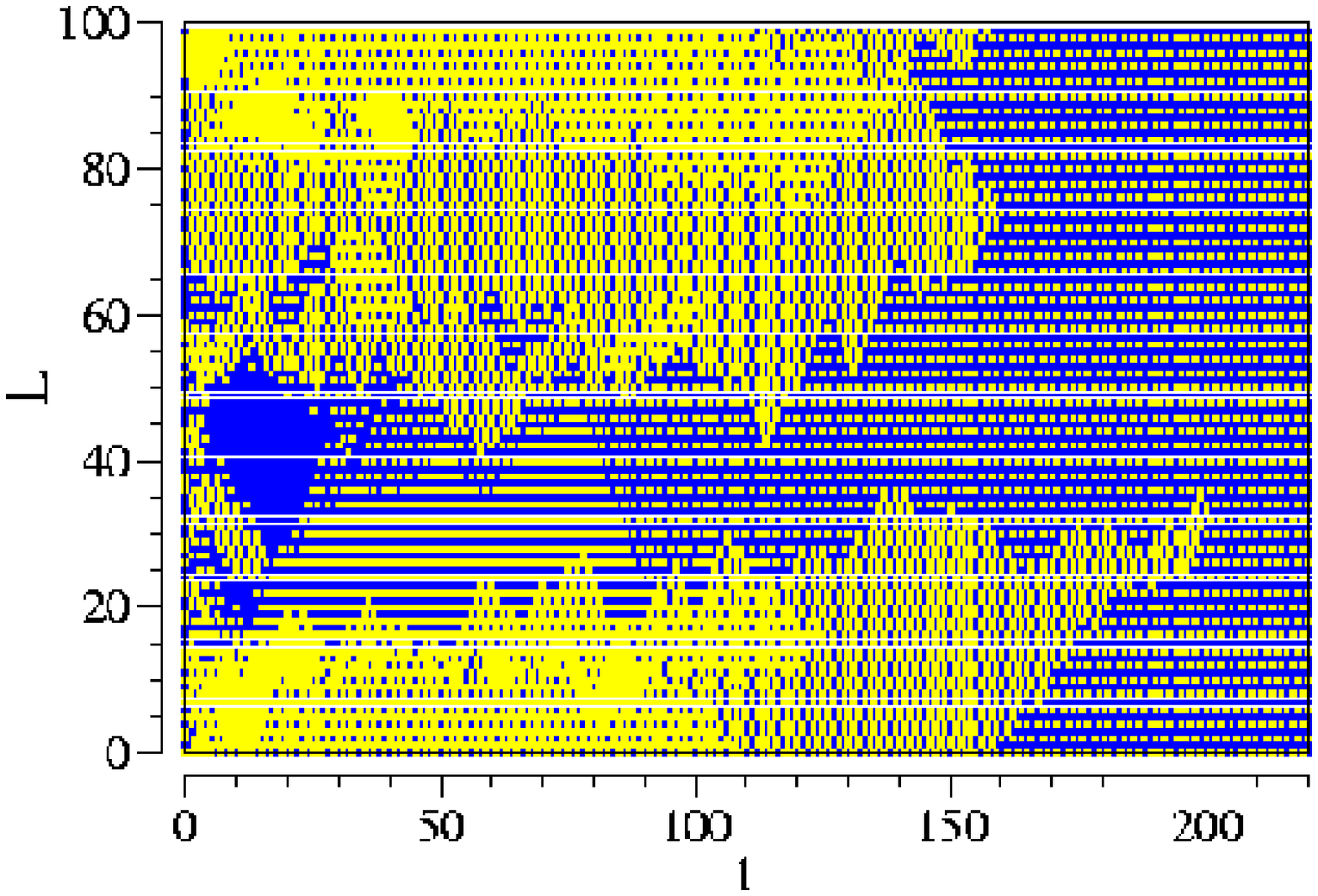,width=6.5cm,angle=-90}\hfill 
\epsfig{file=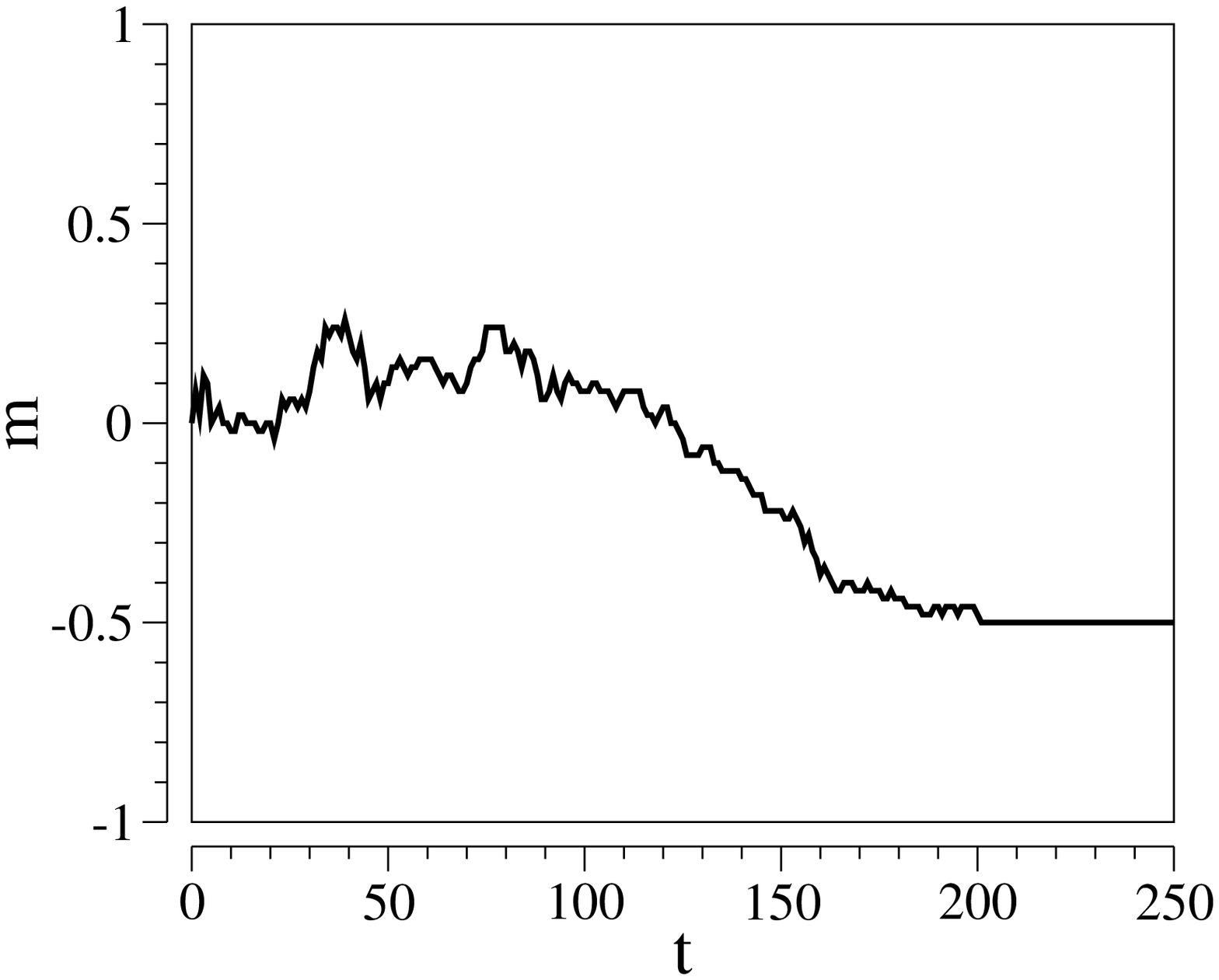,width=6.5cm,angle=-90}
}\vspace*{-10pt}
  \caption{
    Evolution of the one-dimensional lattice (left) and magnetization $m$
    (right) vs time in generations.  Asymptotically, an \emph{asymmetric}
    coexistence attractor is reached.  (Same setup as in
    \pic{fig:Snaid_01}, but \emph{synchronous} update according to the
    voter rules, \eqn{eq:nvoter}, with $\sigma=0.5$)
    \label{fig:coex_as}}
\end{figure}
The influence of the synchronous update rule on the intermediate dynamics
and the stationary distributions is shown in Fig.~\ref{fig:coex_as}. We
see that the spatio-temporal distribution now shows the intermediate
coexistence of six different domains, characterized by the local
configurations given in Table \ref{6attr}. Eventually, attractor 5,
displaying the \emph{asymmetric coexistence} of the opposite opinions, is
reached, which can also be confirmed by looking at the magnetization
$m(t)$.

 \begin{figure}[htbp]
\vspace*{-10pt}
 \centerline{
\epsfig{file=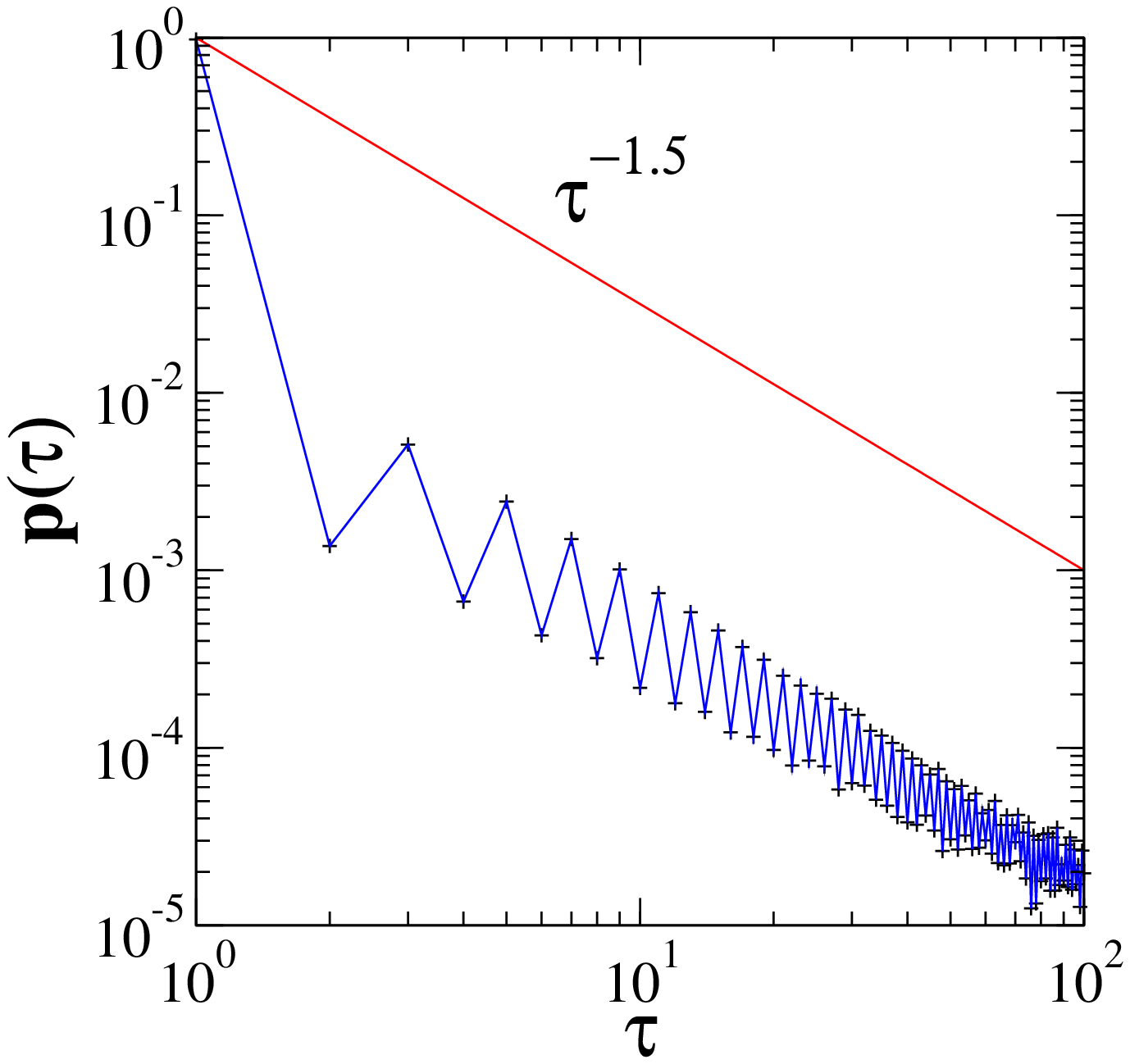,width=8cm}\hfill
\epsfig{file=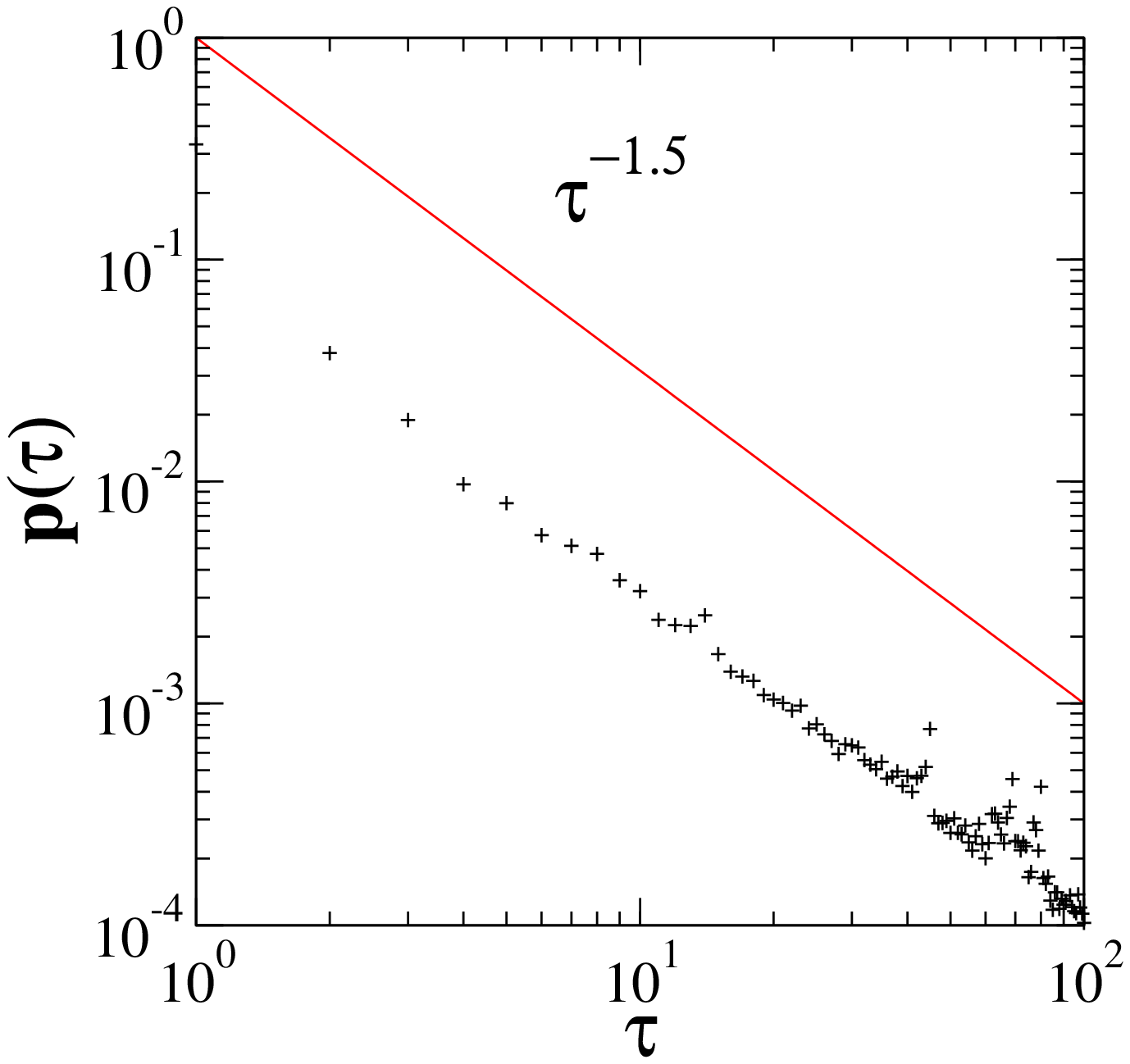,width=8cm}}
\vspace*{-10pt}
   \caption{
     Distribution of decision times $P(\tau)$ averaged over 200
     simulations: (left) $\sigma=0.5$, (right) $\sigma in \{0; 1\}$.
     ($N=100$, synchronous update according to the voter rules,
     \eqn{eq:nvoter})}
   \label{power-sync}
\end{figure}
Finally, we investigate how the synchronous update rule affects the
distribution of decision times, $P(\tau)$. As shown in 
\pic{power-sync}(left), for $\sigma=0.5$ we do \emph{not} find a
power law for the synchronous case. This is due to the fact that during
synchronous update, mostly (i.e. with probability 0.75) domains from the
three new attractors (4-6) appear, even during the intermediate dynamics
(cf.  \pic{fig:coex_as}). Their local configurations, however, force the
``individuals'' to change their opinion during every time step. We can
observe that the mean value of $\tau$ is about 0.93.
However, if $\sigma\in\{0,1\}$ then the three new attractors do not
appear and the known power law $P(\tau)\propto \tau^{-1.5}$ can be
recovered as in the asynchronous case, see \pic{power-sync}(right).

\section{Conclusions}
\label{6}

In this paper we investigate similarities and differences between the
previously established Sznajd model (SM) and the well known voter model
(VM) in one dimension. It is shown that the SM can be completely
reformulated in terms of a \emph{linear} VM, where the transition rates
towards a given opinion are directly proportional to the second-nearest
neighborhood frequency of the respective opinion, \eqn{eq:voter2}.  The
equivalence of the dynamics is demonstrated by extensive computer
simulations that show the same behavior (i) for the spatial-temporal
evolution of the lattice, $L(t)$, $m(t)$, (ii) for the power law
distribution of decision times $P(\tau)$, (iii) for the log-normal
distribution of relaxation times $P(\mu)$, and (iv) for the final
attractor statistics.

We basically conclude that there are no differences between SM and VM
with respect to these indicators. In particular, it does not matter
whether the information flows from inward out (as in SM) or from outward
in (as in VM). Also the fact that in SM dynamics \emph{two} opinions are
changed at the same time, while in VM only \emph{one} opinion is changed,
does not change the dynamic behavior, except that the average time scale
of relaxation is doubled.

So, given that we can reduce the SM dynamics to linear VM dynamics,
what are the advantages of such a reduction? \emph{First}, we could
reveal that in SM only the \emph{second nearest neighbors} matter for the
opinion dynamics, no matter what the nearest neighbors are.
\emph{Second}, in VM we could find a parameter $\sigma$ that expands the
original SM dynamics by considering the case of \emph{frustration}. We
further show that $\sigma$ plays a crucial role in explaining the
\emph{phase transition} known in SM, from the initial random distribution
to either antagonistic or consensus attractors.  \emph{Third}, since the
SM is basically a linear VM, all the techniques developed for VM to
describe the spatial structure formation, e.g. \emph{pair approximations}
of the spatial correlations or Markov chain analysis, can be adapted also
for the analysis of the SM. This will be done in a forthcoming paper.

In this paper, we have also expanded the original SM dynamics by
considering \emph{synchronous} update rules. We show that this will lead
to three additional attractors, which are reached with probability 0.75,
while the original three attractors are reached only with probability
0.25. In the synchronous case, we find a \emph{asymmetric coexistence} of
the different opinions, i.e. the existence of a majority/minority
different from 1/0, which is not found in the original SM.

Finally we address the issue of extending the proposed VM dynamics to a
two-(and higher) dimensional CA. This extension has been done for the SM
already in \citep{sousa:2000}. Also the proposed VM can be easily
extended to two-dimensional problems, based on considerations e.g. in
\citep{fs-voter-03, Molofsky:99}. We just have to adjust the
second-nearest neighbor frequencies to the different neighborhood
definitions (such as Von-Neumann neighborhood with eight, or Moore
neighborhood with sixteen second-nearest neighbors). This shall be also
done in a forthcoming paper.

\subsection*{Acknowledgement}

The authors want to thank Robert Mach for his help in drawing the
spatio-temporal patterns. LB is very much indebted to Heinz
M{\"u}hlenbein for making his stay at Fraunhofer AIS possible. 

\bibliography{jmpc-web}

\end{document}